\documentclass[useAMS,usenatbib]{mn2e}
\usepackage[figuresright]{rotating}
\usepackage{amsmath,amssymb}

\title[Analytical forms of chaotic spiral arms]{Analytical forms of chaotic
spiral arms}
\author[M. Harsoula, C. Efthymiopoulos, and G. Contopoulos]
       {M. ~Harsoula,$^{1}$ \thanks{E-mail: mharsoul@academyofathens.gr}
C. Efthymiopoulos,$^{1}$ \thanks {E-mail:
cefthim@academyofathens.gr}  and G. Contopoulos $^{1}$
\thanks{E-mail:
gcontop@academyofathens.gr}\\
       $^{1}$ Research Center for Astronomy,
           Academy of Athens, Soranou Efesiou 4, GR-115 27 Athens, Greece\\
           }
\begin{document}

\date{ }

\pagerange{\pageref{firstpage}--\pageref{lastpage}} \pubyear{2015}

\maketitle

\label{firstpage}

\begin{abstract}
We develop an analytical theory of chaotic spiral arms in galaxies.
This is based on the Moser theory of invariant manifolds around
unstable periodic orbits. We apply this theory to the chaotic spiral
arms, that start from the neighborhood of the Lagrangian points
$L_1$ and $L_2$ at the end of the bar in a barred-spiral galaxy. The
series representing the invariant manifolds starting at the
Lagrangian points $L_1$, $L_2$, or unstable periodic orbits around
$L_1$ and $L_2$, yield spiral patterns in the configuration space.
These series converge in a domain around every Lagrangian point,
called ``Moser domain'' and represent the orbits that constitute the
chaotic spiral arms. In fact, these orbits are not only along the
invariant manifolds, but also in a domain surrounding the invariant
manifolds. We show further that orbits starting outside the Moser
domain but close to it converge to the boundary of the Moser domain,
which acts as an attractor. These orbits stay for a long time close
to the spiral arms before escaping to infinity.
\end{abstract}


\section{Introduction}
An important development in the theory of nonlinear dynamical
systems was provided by \citet{b2,b3} who proved the {\it
convergence} of the normal form series describing the Hamiltonian
dynamics near an {\it unstable} equilibrium point, or an unstable
periodic orbit. This convergence allows to study {\it chaotic} orbits
by analytical means, i.e. using series. This is in contrast with what
happens in the case of the usual Birkhoff normal form series around
stable invariant points, or stable periodic orbits; it is well known
that these series do not converge, but they are only asymptotic
(see \citet{b1} for a review).

In the present paper we present a connection between Moser's theorem
and the so-called {\it manifold theory} of chaotic spiral arms in
rotating barred galaxies. The manifold theory was proposed in 2006
\citep{b34,b27}  and was explored in detail in a number of
subsequent papers \citep{b28,b31,b35,b36,b37,b17,b4}. The theory
predicts a number of morphological correlations between the spiral
arms and the bar strength and/or the pattern speed (see \citet{b42}
for comparison of these features with observations as well as
\citet{b41} for a review).

The basic element of the manifold theory stems from the form of the
unstable invariant manifolds of the family of short-period Lyapunov
orbits around the unstable Lagrangian equilibria $L_1$ or $L_2$ at
the end of the bar (see section 2). These manifolds, when projected
in the configuration space, take the form of trailing spiral arms.
In the manifold picture, the spiral arms in barred galaxies are
density waves, but, contrary to the case of normal galaxies, they
are composed of chaotic orbits. The backbone of the spiral arms can
be due to the pattern formed by the orbits either all along the
unstable manifolds \citep{b27}, or only at the apsidal positions
along the manifolds \citep{b34}; see \citet{b5}, for a discussion of
the differences between these two models). Furthermore, the chaotic
orbits of the manifold theory can exhibit two distinct behaviors,
i.e., i) they can lead to escapes without recurrences, or ii) they
can have a (possibly quite large) number of recurrences inside and
outside the corotation region. The orbits which exhibit recurrences
belong to a more general chaotic population known as the `hot
population' \citep{b7,b6}. Finally, not only the orbits connected
with $L_1$ or $L_2$, but also those connected to other unstable
periodic orbits in the corotation region may exhibit similar
features and support the chaotic spiral arms \citep{b9,b31}.

Although from a geometrical point of view the invariant manifolds
define spiral patterns, it is a basic fact that their measure is
zero in the entire set of all possible initial conditions in the
chaotic phase space at the corotation region. On the other hand, the
observed spiral arms can only correspond to a non-zero phase space
density of stars. Thus, the question is, how can we build domains of
chaotic orbits, of non-zero measure, around the invariant manifolds.
Our answer in this paper is based on Moser's theorem. Namely, we
will argue below that these domains correspond to the {\it domains
of convergence} of the Moser normal form around the unstable
manifolds.

So far, Moser's theorem was applied in very simple dynamical systems
like mappings \citep{b43,b8}. In simple cases it was
shown that the convergence domain extends to infinity along the
invariant manifolds. Further work on the Moser series allowed us to
find the limits of convergence also away from the invariant
manifolds \citep{b10,b20}. Furthermore, in these simple systems it
was possible to find the Moser domains of convergence of several
unstable periodic orbits. By their overlapping we could find
analytically the heteroclinic points between the various resonances
\citep{b18}. A key result of these studies regards the asymptotic
(in time) behavior of the chaotic orbits with initial conditions
inside or outside a Moser domain of convergence. Namely, we found
that orbits starting outside (but close to) the convergence domain
approach arbitrarily close to the outer limits of this domain
asymptotically in time (although they can never enter inside it).
On the other hand, the chaotic orbits with initial conditions inside
the Moser domain can never exit this domain. In conclusion, the Moser
domain of convergence provides a bounded set of chaotic orbits on non-zero
measure which remain always close to the invariant manifolds, while
the boundary of this domain acts as an {\it attractor} for all the
chaotic initial conditions exterior to the domain (and close enough
to the boundary, see section 3).

In the present paper, we apply the theory of Moser for orbits
starting close to the Lagrangian points $L_1$ and $L_2$. In
particular, we compute the Moser domain of convergence for normal
form series built around the equilibria $L_1$ and $L_2$ in three
different models of barred galaxies emerging from past numerical
simulations \citep{b33}. This allows to obtain analytically not only
the form of the invariant manifolds, which define the spiral arms,
but also the form of the Moser domain of convergence. Then, we show
that this domain follows closely the spiral patterns, and provides
a chaotic set of non-zero measure along the spiral patterns.

The paper is structured as follows: section 2 briefly presents  a
summary of the manifold theory and the models used in the present
paper. Section 3 presents the normal form analytical computations,
the computation of the Moser domain of convergence, based on
high-order series expansions carried by a computer-algebraic
program, and the results, which illustrate the connection between
Moser domains and spiral patterns. In section 4 we provide a
theoretical interpretation based on an approximative simplified
mapping model. Finally, section 5 summarizes our basic conclusions.

\section{Manifold theory and models}

\subsection{Manifold theory}

The Hamiltonian of motion in the plane of a galaxy with a rotating
bar for a test particle of mass $m=1$ is given by:
\begin{equation}\label{hamrot}
H={1\over 2} \left(p_x^2+p_y^2\right)-\Omega_p(xp_y-yp_x) +\Phi(x,y)
\end{equation}
where $x,y$ are the Cartesian positions in the rotating frame with
pattern speed $\Omega_p$, $p_x,p_y$ are the canonical momenta
(velocities) in an instantaneous rest frame with axes ($x,y$), and
$\Phi(x,y)$ is the gravitational potential of the galaxy in the
rotating frame. For simplicity, we consider the potential as
time-independent.

The unstable Lagrangian point $L_1$ (and similarly $L_2$)
corresponds to a solution
$(x,y,p_x,p_y)=(x_{L_1},y_{L_1},p_{xL_1},p_{yL_1})$ of the
equilibrium equations $\partial{H}/\partial x$ $=$
$\partial{H}/\partial y$ $=$ $\partial{H}/\partial p_x$ $=$
$\partial{H}/\partial p_y$ $=0$. The equilibrium points $L_1$ and
$L_2$ are located at the end of the bar, and they are simply
unstable, i.e. the matrix of linearized equations around each of
these points has two imaginary and two real eigenvalues
$\lambda_{1,2}=\pm i\omega_0$, $\lambda_{3,4}=\pm\nu_0$, with
$\omega_0,\nu_0$ real. Then, we can introduce a symplectic change
of variables $(x,y,p_x,p_y)$ $\rightarrow$ $(q,u,p,v)$, where
($q,p$) and ($u,v$) are conjugate pairs such that in the new
variables the Hamiltonian can be expanded around $L_1$ (or $L_2$)
as:
\begin{equation}\label{hamlinl1}
H=\omega_0\left({q^2+p^2\over 2}\right) +\nu_0 u v +
\sum_{s=3}^{\infty} P_s(q,p,u,v)
\end{equation}
where the $P_s$ are polynomial functions of degree $s$. Let us
neglect, to the lowest order limit, the effect of the terms $P_s$.
Then, the motion in the $(q,p)$ plane reduces to the limit of a
harmonic oscillation with frequency $\omega_0$, and the plane
$(q,p)$ is called the linear center manifold of the equilibrium
point $L_1$. On the other hand, the variable $u$ grows
exponentially, $u=u_0 e^{\nu_0 t}$, while the variable $v$ decays
exponentially as $v=v_0 e^{-\nu_0 t}$. Then, the $u$ axis defines
the linear unstable manifold, and the $v$ axis the linear stable
manifold of the equilibrium point $L_1$.

Back-transforming to the original cartesian variables, $(q,u,p,v)$
$\rightarrow$ $(x,y,p_x,p_y)$, the independent motions in the
$(q,p)$ plane and in the $u$-axis, and $v$-axis lead to the following:

i) The oscillations in $(q,p)$ define retrograde epicyclic motions
around $L_1$ with frequency $\omega_0$. In the full nonlinear
problem, these are continued as a family of retrograde periodic
orbits of period $2\pi/\omega$, around $L_1$, called the `short
period family of orbits '$PL_1$' \citep{b34} or the horizontal
`Lyapunov family of orbits' \citep{b27}. In general we have
$\omega\simeq\omega_0$, with the difference $\omega-\omega_0$
increasing with the size of the epicycle.

ii) The variable $u$ grows in the forward sense of time
($t\rightarrow\infty$). In Cartesian variables, this growth
describes a recession of the guiding center of the epicycle away
from $L_1$ in the trailing sense \citep [see e.g. Fig.1 of] []{b35}.
Then, the combined guiding center and epicyclic motion forms a tube
in the plane $(x,y)$. This tube corresponds to a two-dimensional
surface in the full phase-space (including the velocities), and it
is called the unstable invariant manifold of the periodic orbit
$PL_1$. Hereafter, it is denoted by $W_{PL_1}^U$. Likewise, the
variable $v$ grows in the reverse sense of time
$t\rightarrow-\infty$. In this case, the corresponding tube forms
the stable manifold of $PL_1$ (denoted hereafter by $W_{PL_1}^S$).
In the forward sense of time, every initial condition on
$W_{PL_1}^S$ leads to an orbit tending asymptotically closer and
closer towards the periodic orbit $PL_1$.

Depending, now, on the parameters of the galactic model (e.g the bar
strength and/or the pattern speed), at large distances from the
points $L_1$ and $L_2$ we distinguish two cases: i) the orbits along
the manifolds $W_{PL_1}^U$, $W_{PL_2}^U$ are led directly to
escapes, or ii) the orbits become, at least temporarily, chaotically
recurrent. In the latter case, the orbits (and hence the patterns
formed by the invariant manifolds) make several oscillations inside
and outside corotation. Then, the tubes of the invariant manifolds
exhibit an intricate shape which is no longer a simple spiral (see,
for example Fig.12 of \citet{b31}, or Fig.1  of \citet{b4}).
However, in \citet{b33} is was shown that if we only consider the
locus of all points on the manifolds where the orbits come to an
{\it apocentric} position, this locus still has the form of trailing
spiral arms. On the other hand, the locus of pericentric manifold
positions takes the form of either the outer envelope of the bar, or
the innermost part of the spiral arms which can, sometimes, have a
shape of a ring.

The galactic models studied in the present paper correspond all to
the case (ii) above. In the sequel we first illustrate the mechanism
of generation of spiral arms by the apocentric loci of the invariant
manifolds in three rotating barred-galaxy models produced by N-Body
simulations \citep{b33}. These models are summarized in the next
subsection. Then, in the next section we employ them as examples in
order to demonstrate our present new result, i.e., the connection
between Moser domains and spiral structure.

\subsection{Models}

In our numerical demonstrations below we use the same N-body models
as in \citet{b33}, called there the experiments $QR2$, $QR3$ and
$QR4$. We call them models "A", "B" and "C" respectively, hereafter.
The various features, approximations, and limitations of these
simulations are discussed in detail in \citet{b33} and \citet{b31}
(see also Appendix A). Here, we are only interested in some
characteristic snapshots in each simulation, in which the simulation
exhibits a conspicuous bi-symmetric spiral structure (typically, in
this type of simulations, the spiral structure appears and
disappears recurrently in time, see \citet{b7}). Namely, after
choosing one such snapshot, we extract the instantaneous N-body
potential and thereby consider a frozen in time potential model.
Likewise, we extract the instantaneous value of the pattern speed.
This allows to numerically define a 2D Hamiltonian for the orbits in
the disc plane, which in polar coordinates is given by:
\begin{equation}\label{hamrot2}
H= \frac{P_r^2}{2}+\frac{P_{\phi}^2}{2r^2}-\Omega_p ~P_{\phi}
+\Phi(r,\phi)
\end{equation}
In this expression, $(r, \phi)$ are polar coordinates in the
rotating frame, $P_r = \dot{r}$ and $P_{\phi} = r^2(\dot{\phi} +
\Omega_p)$ is the angular momentum in the rest frame.

\begin{figure}
\centering
\includegraphics[scale=0.30]{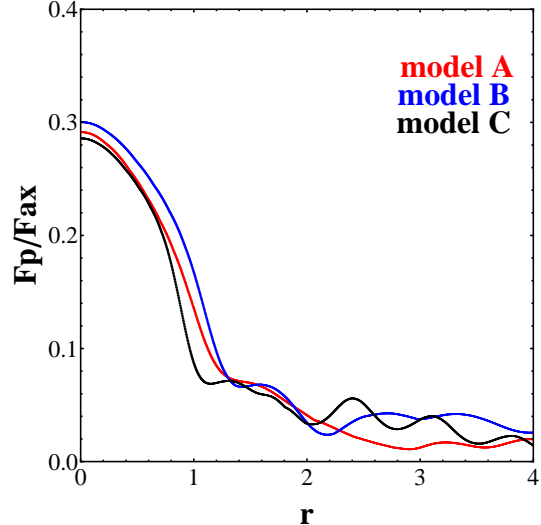}
\caption{The ratio $Q=F_p/F_{ax}$ as a function of $r$, where
$F_p$ is the maximum (with respect to the azimuth $\phi$) absolute
value of the transverse force at a given cylindrical radius $r$ in
the disc plane, and $F_{ax}$ is the mean absolute value (over
$\phi$) of the radial force at the same radius $r$. A non-zero
$Q-$ value measures the strength of the non-axisymmetric
perturbation of the bar and of the spiral structure. The three
curves correspond to models  A (red, light gray in printed
version), B (blue, gray in printed version), and C (black). The
outer oscillations of the three curves are due to the spiral
perturbation. } \label{force}
\end{figure}

To simplify computations, we only consider the $m=2$ mode of the
galactic bar. Then, the potential $\Phi(r,\phi)$ in our galactic
models is given by:
\begin{equation}\label{phipot}
\Phi(r,\phi)=\Phi_0(r)+\Phi_1(r)\cos 2\phi+\Phi_2(r)\sin 2\phi
\end{equation}
where $\Phi_0(r)$ is the axisymmetric potential while the second and
third terms of Eq.(\ref{phipot}) correspond to an $m=2$ mode of the
non-axisymmetric potential perturbation. The formulas for
$\Phi_0(r)$, $\Phi_1(r)$ and $\Phi_2(r)$ are derived following
\citet{allen90} and they read:
\begin{eqnarray}
\Phi_0(r)=-\frac{1}{R}(A_{00}+\frac{1}{4}A_{20}-\frac{3}{2}A_{22})\nonumber
\\\Phi_1(r)= -\frac{3}{2R}(\frac{1}{2}A_{20}+A_{22})
\\\Phi_2(r)=+\frac{3}{2R}A_{21}\nonumber
\end{eqnarray}
with
\begin{eqnarray}\label{coef}
A_{00}=\displaystyle\sum_{n=0}^{19}[B_{n00}.j_0(\xi_{n0})]\nonumber
\\A_{20}=\displaystyle\sum_{n=0}^{19}[B_{n20}.j_2(\xi_{n2})]
\\A_{21}=\displaystyle\sum_{n=0}^{19}[C_{n21}.j_2(\xi_{n2})]\nonumber
\\A_{22}=\displaystyle\sum_{n=0}^{19}[B_{n22}.j_2(\xi_{n2})]\nonumber
\end{eqnarray}
where $n$ is the number of the so-called `radial' terms in the
series expansions of the potential, $j_0,~j_2$ are spherical Bessel
functions (see \citet{b38}) and $\xi_{nl}=a_{nl}r/R$, where $R$ is a
numerical constant representing the size of the system (defined as
the N-body code boundary where the solutions of the Poisson equation
are matched with the solutions of the Laplace equation, see
\citet{b31}). Finally $a_{n0}$ are the roots of the equation
$j_{n-1}(a)=0$, yielding $a_{n0}=(n+1/2)\pi$, while $a_{n2}$ are the
roots of the equation $\tan(a_{n2})=a_{n2})$ (the latter are found
numerically). The coefficients $B_{n00},~B_{n20},~C_{n21},~B_{n22}$
are calculated by the N-body code using the positions of the N-body
particles and the Poisson equation $\nabla^2\Phi=4\pi G\rho$, where
$\rho$ is the density of matter. We provide plots of the
coefficients for all three galactic models in the Appendix A. The
numerical values of the coefficients can be provided to any one
interested after private communication.

\begin{figure*}
\centering
\includegraphics[scale=0.26]{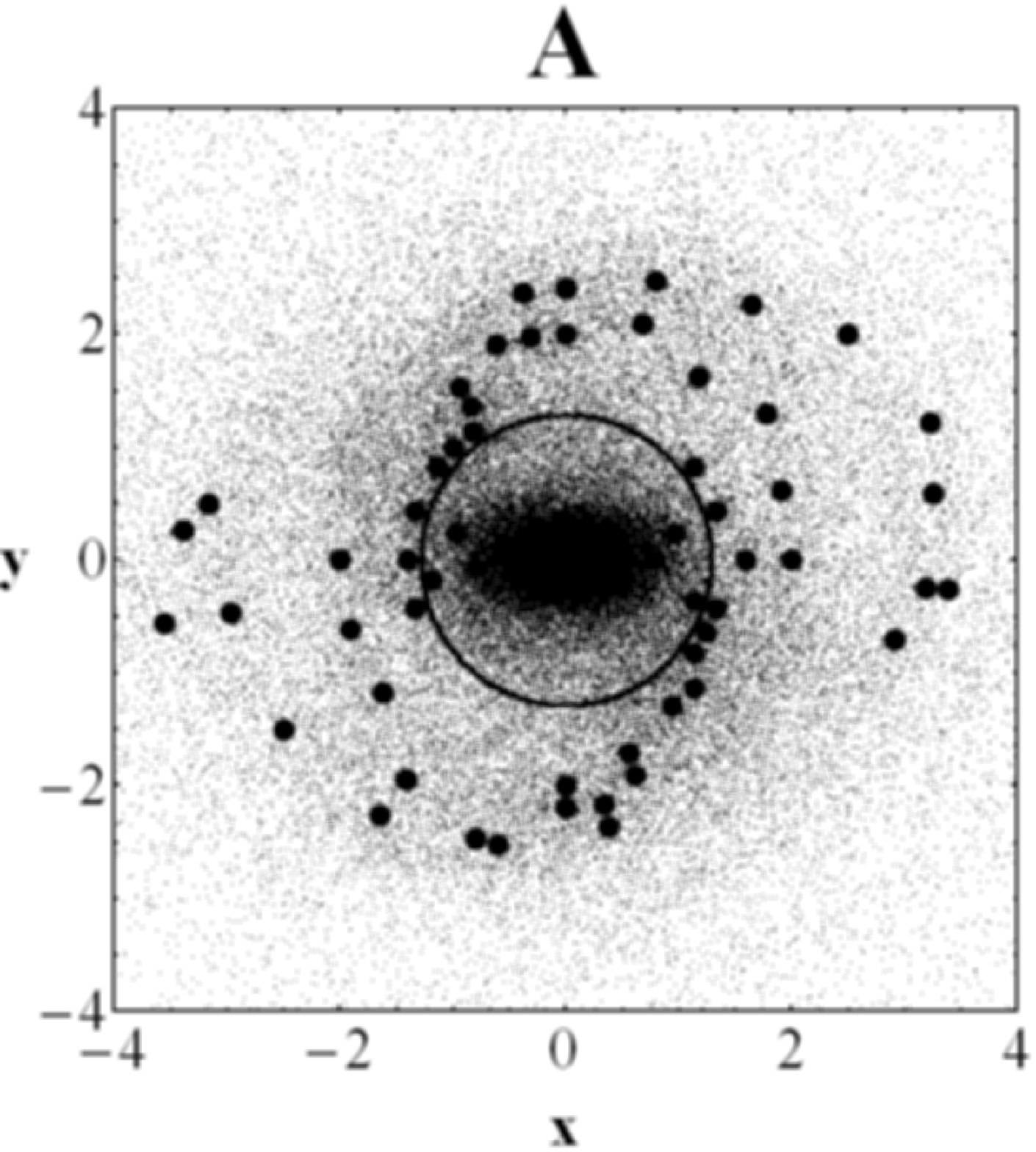}
\includegraphics[scale=0.26]{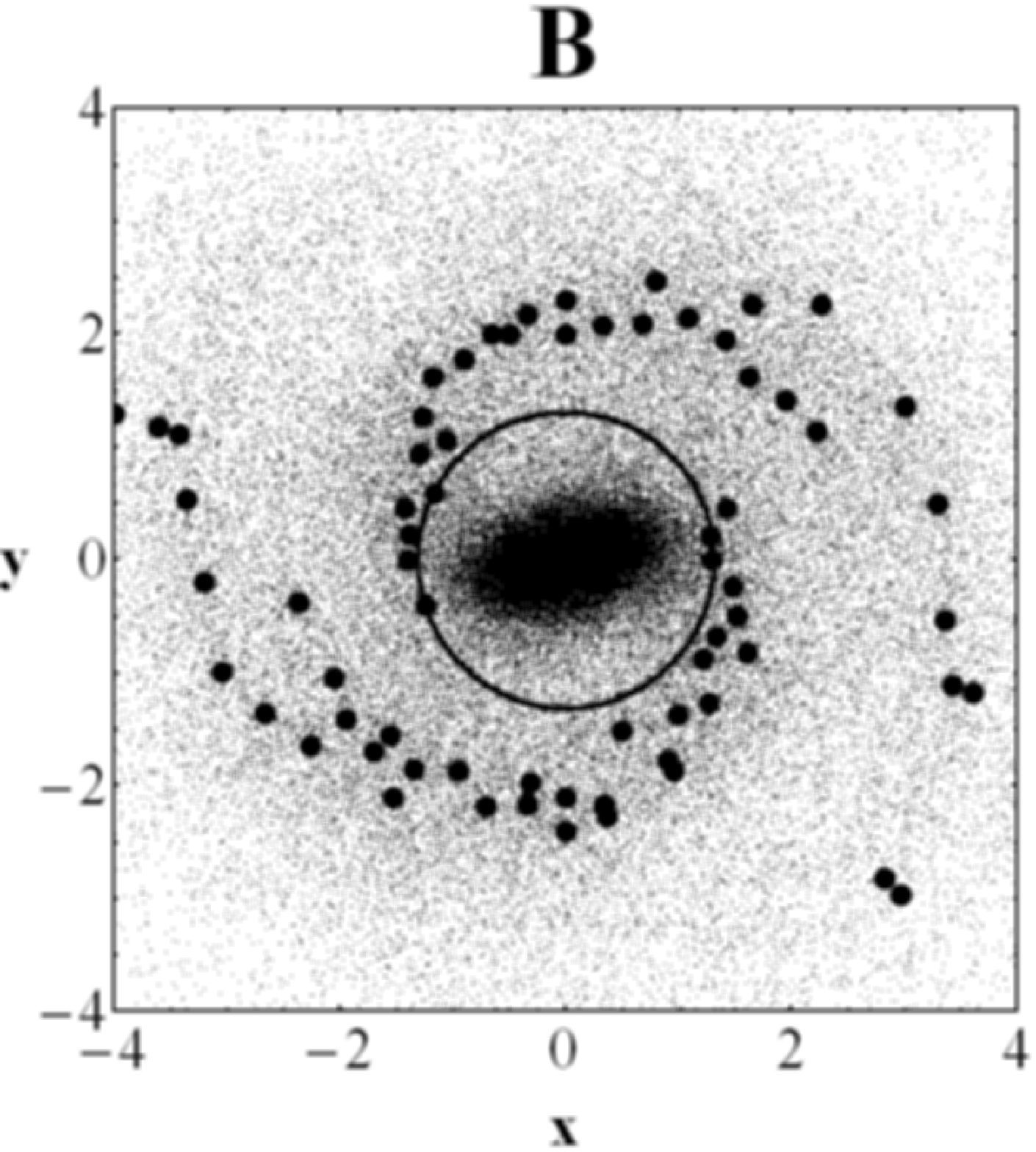}
\includegraphics[scale=0.26]{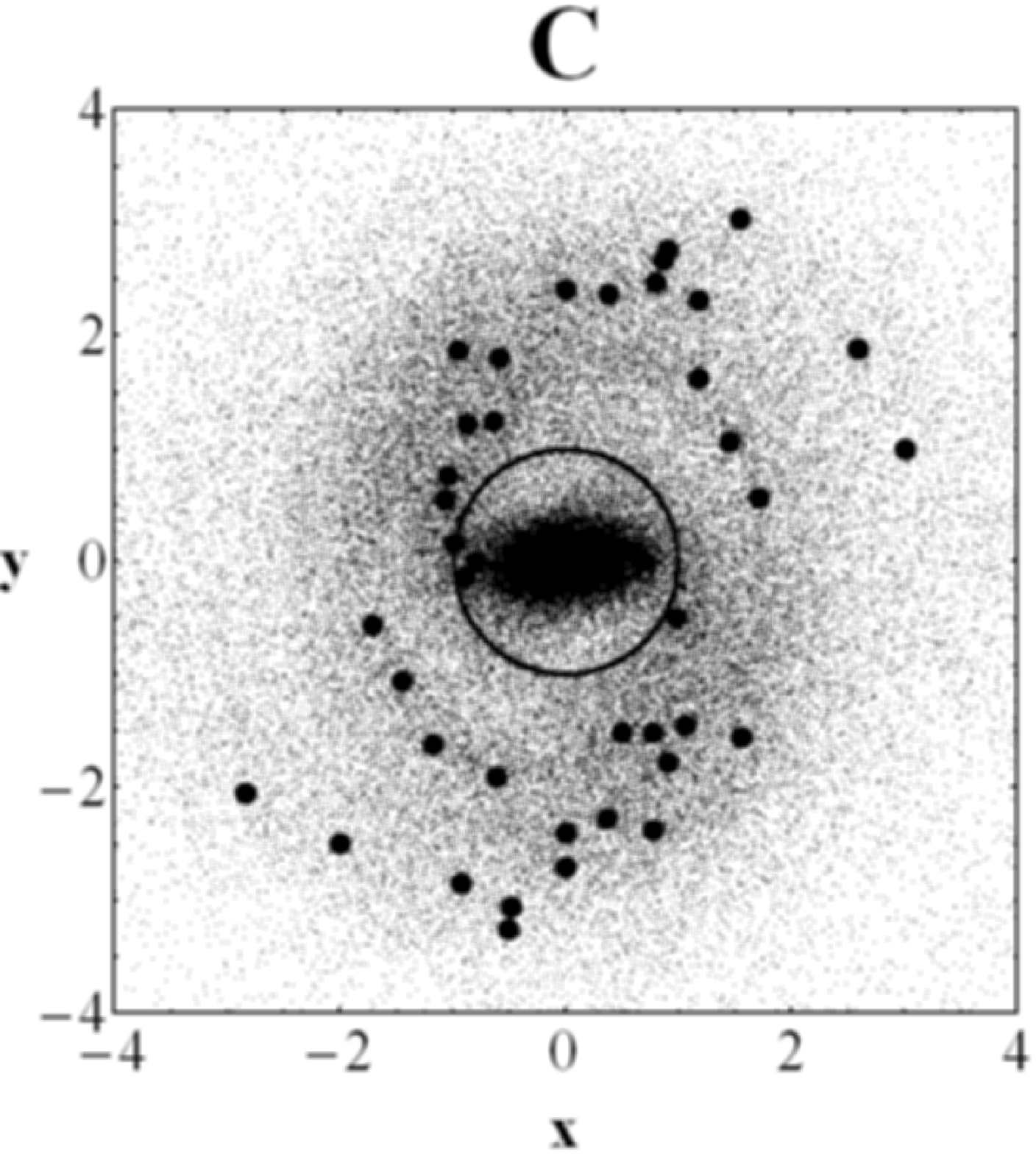}
\includegraphics[scale=0.45]{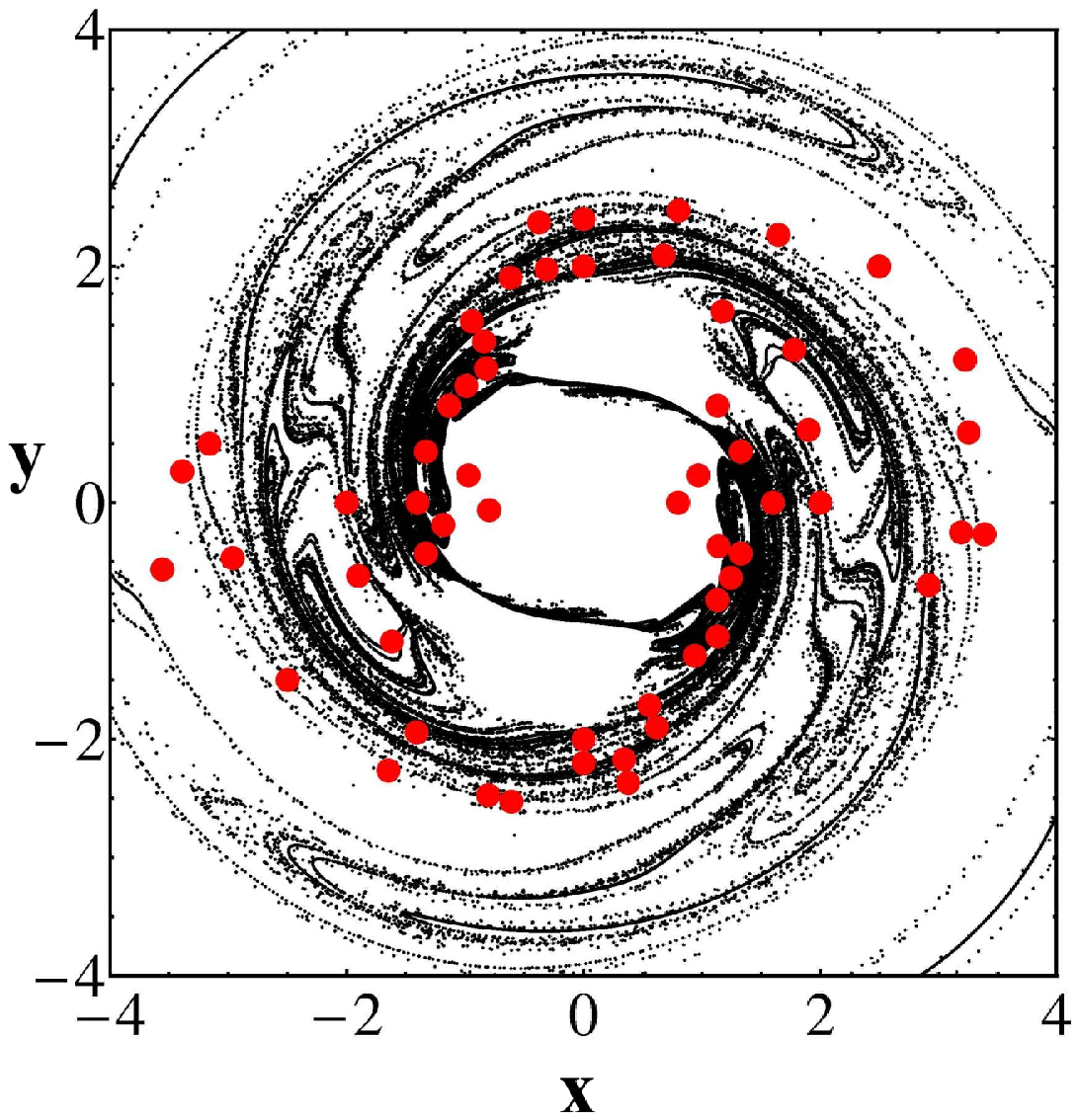}
\includegraphics[scale=0.45]{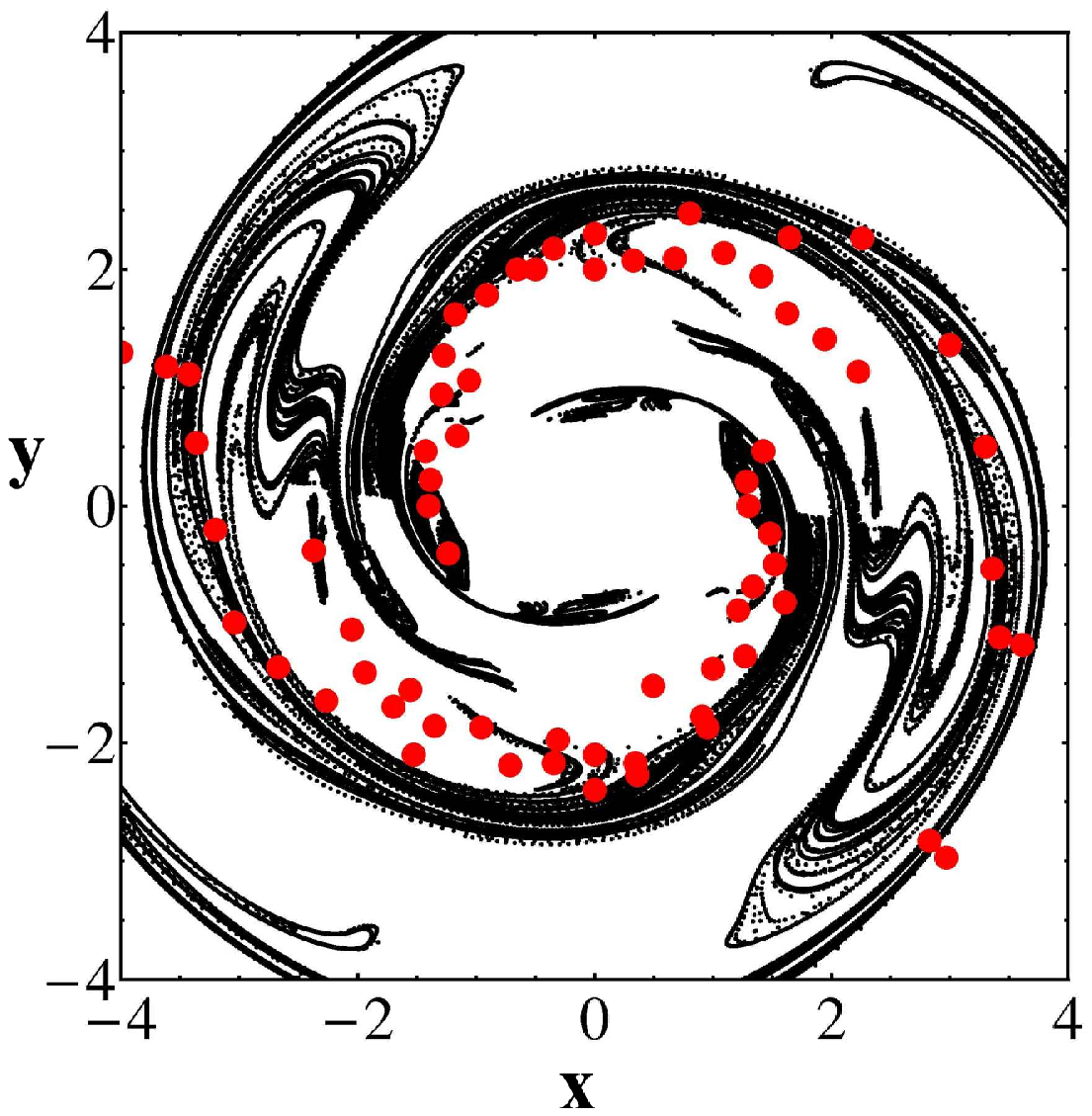}
\includegraphics[scale=0.45]{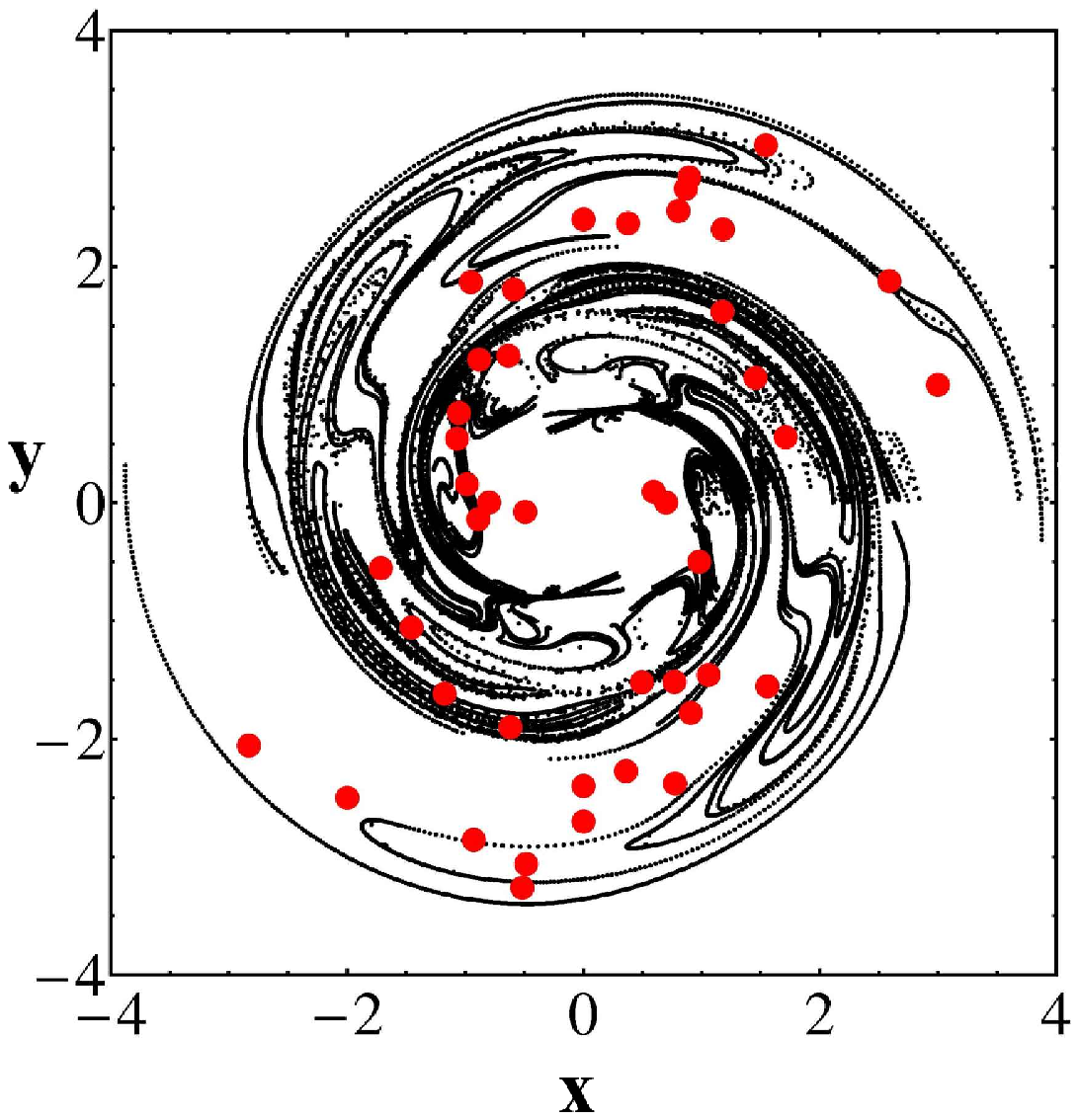}
\caption{\textbf{\textit{First line}}: Projection on the plane of
rotation of the three N-body galactic models, "A", "B" and "C".
The circle of corotation is superimposed. The thick dots show the
local maxima of the projected surface density.
\textbf{\textit{Second line}}: The numerical apocentric manifolds
that track the maxima of the density for the three galactic models
"A", "B" and "C" respectively. The red thick dots are the same as
the black thick dots of the upper line.} \label{bqr2}
\end{figure*}

In Figure \ref{force} we plot the perturbation of each galactic
model, given by the maximum ratio of the non-axisymmetric forces,
due to the bar and to the spiral arms, versus total axisymmetric
force, as a function of the distance $r$ for models A (red), B (blue)
and C (black).

The units used in the present paper are the N-body code units. The
relation of these units to natural units is discussed in detail in
\citet{b31}. As a rough guide to figures 2,4,5,6 below, one unit of
length corresponds to $\approx8$~Kpc.

\subsection{Numerical invariant manifolds}
In Fig.\ref{bqr2}, in the first line  we plot the three N-body
galactic models at the snapshot at which we make the analysis. The
circle of corotation is superimposed. The thick dots correspond to
the local maxima of the surface density $\Sigma(r,\phi)$ of the
N-body particles in the simulation. We find, in general, one
prominent maximum along each direction and some secondary maxima.
 The prominent maxima always define a conspicuous spiral pattern. The secondary maxima are also observed to from patterns like spiral arms or rings.
  In the second line of Fig.\ref{bqr2}, now, we plot the apocentric manifolds of the
$PL_1$ and $PL_2$ orbits for a Jacobi constant close to the one at
$L_1$ (differing from it at the fourth digit). In fact, the plot
consists of the first two apocenters of orbits having initial
conditions along the unstable asymptotic manifold of the $PL_1$
and $PL_2$ orbits at an energy level close to the one that
corresponds to the Lagrangian point $L_1$.  Some of these results
have been published already in past papers \citep[see][]{b34,b31}
and consistently demonstrate a fact of key importance, namely that
a large fraction of chaotic orbits outside corotation have many
recurrences along spiral segments, i.e. they exhibit stickiness
effects. Therefore they can support the spiral structure of the
galaxy for considerably long times of the order of several decades
of galactic periods before escaping from the system
\citep{b17,Har2013}.

In the sequel we will derive similar figures analytically, using
convergent series of the normal form of the Hamiltonian for the
three galactic models. We will thus establish a connection between
the domain of the spiral structure and the Moser domain of
convergence (see section 3.2).

\section{Analytical description of the spiral arms region}
In the present section, we first give the algorithm of computation
of the Moser normal form around an unstable equilibrium point. We
also introduce some relevant notation and terminology. Then, we
implement the method in order to analytically determine the Moser
domains, as well as the thereby induced loci of the spiral arms in
our specific galactic models described in the previous section.

\subsection{``Moser'' normal form construction}
{\it (i) Hamiltonian expansion.}
The first step for the normal form construction is the expansion of
the Hamiltonian (\ref{hamrot2}) around the Lagrangian point $L_1$.
Hamilton's equations yield the exact position of the five Lagrangian
points as the stationary points of the effective potential
$\Phi_{eff}=\Phi(r,\phi)-\Omega_p P_{\phi}$, i.e. the solutions
of the following equations:
\begin{eqnarray}
\frac{dr}{dt}=\frac{\partial H}{\partial P_r}=0,~~
\frac{dP_r}{dt}=-\frac{\partial H}{\partial r}=0  \nonumber\\
 \frac{d \phi}{dt}=\frac{\partial H}{\partial P_{\phi}}=0,~~
\frac{d P_{\phi}}{dt}=-\frac{\partial H}{\partial \phi}=0
\end{eqnarray}
Let $(r_{L_1},~P_{r_{L_1}},~\phi_{L_1},~P_{{\phi}_{L_1}})$ denote
the solution for the Lagrangian point $L_1$. The radius $r_{L_1}$ is
close to the corotation radius $r_c$, while the angle $\phi_{L_1}$
is in a direction close to the bar's major axis. We expand the
Hamiltonian (\ref{hamrot2}) in series around ($r_{L_1},\phi_{L_1},
P_{r_{L_1}},P_{{\phi}_{L_1}}$) by making the following
substitutions:
\begin{equation}
r\rightarrow r_{L_1}+ \delta r,~~P_r \rightarrow
P_{r_{L_1}}+P_x,~~\phi \rightarrow \phi_{L_1}+\delta \phi,~~P_{\phi}
\rightarrow P_{{\phi}_{L_1}}+J_{\phi}
\end{equation}
Since the expansion is around an equilibrium point it contains no
terms linear in $(\delta r,\delta\phi,P_x,J_\phi)$. On the other
hand, the quadratic terms $H_2(\delta r,\delta\phi,P_x,J_{\phi})$
yield the linearized equations of motion around the equilibrium
which are of the form
\begin{equation}\label{eqvar}
 \left( \begin{array}{c}
  \dot{\delta r}  \\
\dot{\delta \phi}\\
\dot{ P_x}\\
  \dot{ J_{\phi}} \end{array} \right)
=M\left( \begin{array}{c}
  \delta r  \\
\delta \phi\\
 P_x\\
 J_{\phi} \end{array} \right)
\end{equation}
or $\dot{X}=M X$ where $X=(\delta r,\delta\phi,P_x,J_\phi)^T$ and
$M$ is the $4\times 4$ characteristic variational matrix with
constant coefficients.

We now introduce a linear transformation $X=A\cdot U$ rendering the
linear system (\ref{eqvar}) diagonal in a set of new canonical
variables $U\equiv (\alpha,u,\beta,v)^T$. We require that in the new
variables the linearized equations take the form
$\dot{U}=\Lambda\cdot U$ where $\Lambda$ is the $4\times4$ matrix:
\begin{equation}\label{lammat}
\Lambda= \left( \begin{array}{cccc}
  \lambda_1 & 0&0&0  \\
0&  \lambda_2&0&0\\
 0&0&\lambda_3&0\\
 0&0&0& \lambda_4 \end{array} \right)
\end{equation}
with $\lambda_1=i\omega_0$, $\lambda_2=-i\omega_0$,
$\lambda_3=\nu_0$, $\lambda_4=-\nu_0$, being the four eigenvalues of
the matrix $M$. We note that since $L_1$ is simply unstable it
necessarily has a pair of opposite imaginary and a pair of opposite
real eigenvalues. It is easy to show that the above requirements
imply that the matrix $A$ contains as columns four linearly
independent eigenvectors of the variational matrix $M$ corresponding
to the eigevalues $\lambda_i$, $i=1,\ldots,4$ respectively. Each of
these eigenvectors can be specified by solving the characteristic
system $M\cdot A = A\cdot\Lambda$. Since this system is homogeneous
the solution for each eigenvector is specified up to an arbitrary
multiplicative constant. We exploit this arbitrariness in order to
render the linear transformation $X\rightarrow U$ symplectic. To
this end, starting from any initial solution $A$, we define a new
matrix $\overline{A}$ by multiplying the first and third columns of $A$
by an unspecified coefficient $c_1$, and the second and fourth columns
by an unspecified coefficient $c_2$. Finally, we specify the values of
$c_1$ and $c_2$ by the requirement that the condition
\begin{equation}\label{sympl}
\overline{A}~{\cal J}_4~\overline{A}^T={\cal J}_4
\end{equation}
be satisfied, where ${\cal J}_4$ is the $4\times 4$ fundamental
symplectic matrix
\begin{equation}
{\cal J}_4=
\begin{bmatrix}
    0       & I_2 \\
    -I_2    & 0 \\
\end{bmatrix}
\end{equation}
where $I_2$ is the $2\times 2$ identity matrix. This determines the
final symplectic transformation $X=\overline{A}\cdot U$.

Once the new matrix $\overline{A}$ is found we can write the
expanded Hamiltonian $H$ as function of the new variables
($\alpha,u,\beta,v$), where ($\alpha,\beta$) and ($u,v$) are
canonically conjugate pairs. The Hamiltonian acquires now a
polynomial form:
\begin{equation}\label{hamser}
H= i ~\omega_0~\alpha~\beta + \nu_0~u~v + \sum_{s=3}^{N_t}
P_s(\alpha,u,\beta,v)
\end{equation}
where
\begin{equation}\label{hamserpol}
P_s(\alpha,u,\beta,v) = \sum_{\stackrel{k_1,k_2,l_1,l_2\geq
0,}{k_1+k_2+l_1+l_2=s}} {\cal A}_{k_1,k_2,l_1,l_2} \alpha^{k_1}
u^{k_2} \beta^{l_1} v^{l_2} \end{equation}
 are polynomials of degree
$s$ with constant coefficients ${\cal A}_{k_1,k_2,l_1,l_2}$, and
$N_t$ is an (inevitably finite in the computer) truncation order. In
all subsequent computations we set $N_t=20$, having checked that
such an order is sufficient to accurately represent the Hamiltonian
expansion in a domain of size $\sim r_c$ (the corotation radius)
around the Lagrangian point $L_1$ with errors of order $10^{-8}$.

It is easy to see that the Hamiltonian (\ref{hamser}) is of the
general form (\ref{hamlinl1}), after the linear symplectic
transformation $\alpha=(q-ip)/\sqrt{2}$, $\beta=(iq-p)/\sqrt{2}$.
Thus, the complex canonical variables $(\alpha,\beta)$ still
represent harmonic oscillator variables (they are known as the
`Birkhoff variables'). Their relation to the harmonic oscillator
action-angle variables $(J,\theta)$ is
$\alpha=-i\sqrt{J}e^{i\theta}$, $\beta=\sqrt{J}e^{-i\theta}$.
Then $J=i\alpha\beta$ is an integral
of the linearized equations of motion. \\
\\
\noindent{\it (ii) Hamiltonian normalization}. We now introduce
a symplectic transformation of the variables $(\alpha,u,\beta,v)$
of the general form
\begin{eqnarray}\label{phitra}
\alpha&=&\Phi_{\alpha}(a,\xi,b,\eta)\nonumber\\
u&=&\Phi_{u}(a,\xi,b,\eta)\\
\beta&=&\Phi_{\beta}(a,\xi,b,\eta)\nonumber\\
v&=&\Phi_{v}(a,\xi,b,\eta)\nonumber
\end{eqnarray}
aiming to render separable the Hamiltonian (\ref{hamser}) in the
new variables $(a,\xi,b,\eta)$, where ($a,b$) and ($\xi,\eta$) are
canonically conjugate pairs. In particular, Moser's theorem
guarantees that there is a transformation of the form
(\ref{phitra}), in which the functions $\Phi_{\alpha}$,
$\Phi_{u}$, $\Phi_{\beta}$, $\Phi_{v}$ are given as {\it
convergent} polynomial series in a domain of the space of the new
variables $(a,\xi,b,\eta)$ surrounding the origin, such that, in
the new variables, the Hamiltonian becomes a function of only the
products $I=iab$, $c=\xi\eta$. This new expression of the
Hamiltonian $H=Z(I,c)$ is called hereafter the `Moser normal
form'.

The details of the algebraic procedure by which we determine the
transformation series (\ref{phitra}) are described in detail in
\citep{b39} (see also \citet{b45}, section 2.10, for a
tutorial). Here we only summarize the formulas implementing the
algorithm in the computer. Let $N_t$ be the maximum truncation
order. The transformation for the variable $\alpha$ (and similarly
for all three remaining variables in Eq.(\ref{phitra}) is given by:
\begin{equation}\label{phia}
\alpha=\exp(L_{\chi_{Nt}})\exp(L_{\chi_{Nt-1}})...
\exp(L_{\chi_2})\exp(L_{\chi_1}) a
\end{equation}
where the quantities $\chi_r$, $r=1,2,\ldots,N_t$, called the "Lie
generating functions", are polynomial functions of degree $r+2$ in
the variables $(a,\xi,b,\eta)$. The symbol $\exp(L_{\chi_r})$
denotes the exponential Lie operator
\begin{equation}\label{lieop}
\exp(L_{\chi_r})=1 + L_{\chi_r}+{1\over 2}L_{\chi_r}^2 +\ldots =
\sum_{n=0}^{\infty}{1\over n!}L^n_{\chi_r}
\end{equation}
where $L_{\chi_r}$ is the Poisson bracket operator, defined for any
function $g(a,\xi,b,\eta)$ by
\begin{equation}\label{poiss}
L_{\chi_r} g = {\partial g\over \partial a} {\partial \chi_r\over
\partial b} +{\partial g\over \partial \xi} {\partial \chi_r\over
\partial \eta} -{\partial g\over \partial b} {\partial \chi_r\over
\partial a} -{\partial g\over \partial \eta} {\partial \chi_r\over
\partial \xi}
\end{equation}
In summary, the transformations (\ref{phitra}) are found by a
sequence of Poisson bracket operations on the variables
$(a,\xi,b,\eta)$, which make use of certain generating functions
$\chi_r$, specified in an appropriate way explained just below. In
the computer, we truncate any repeated Poisson bracket operation at
the point where the operation starts yielding terms of degree higher
than $N_t+2$. This yields eventually a finite truncation of each of
the four series of Eq.(\ref{phitra}).

The functions $\chi_r$, now, are specified step by step by a
recursive algorithm. Let $H^{(0)}(a,\xi,b,\eta)\equiv
H(a,\xi,b,\eta)$. Assume that $r$ steps of the algorithm were
completed. The $r^{th+1}$
 function $\chi_{r+1}$ is the solution of
the equation
\begin{equation}\label{homol}
\{i\omega_0 ab+\nu_0 \xi\eta,\chi_{r+1}\}+h^{(r)}_{r+1}=0
\end{equation}
where $h^{(r)}_{r+1}$ is a polynomial function of degree
$r+3$ which contains all the monomial terms of the function
\begin{equation}\label{newhamil}
H^{(r)}= \exp(L_{\chi_{r}})\exp(L_{\chi_{r-1}})...
\exp(L_{\chi_2})\exp(L_{\chi_1})H^{(0)}
\end{equation}
which are of form
$c_{k_1,k_2,l_1,l_2}a^{k_1}\xi^{k_2}b^{l_1}\eta^{l_2}$ such that
$k_1+k_2+l_1+l_2=r+2$ and $k_1\neq l_1$ or $k_2\neq l_2$. The
solution of Eq.(\ref{homol}) is straightforward. Namely, the
solution is found by the rule:
$$
\mbox{For every term~}
c_{k_1,k_2,l_1,l_2}a^{k_1}\xi^{k_2}b^{l_1}\eta^{l_2}
\mbox{~in~}h^{(r)}_{r+1}
$$
$$
\mbox{add the term~}
{c_{k_1,k_2,l_1,l_2}a^{k_1}\xi^{k_2}b^{l_1}\eta^{l_2} \over
i(k_1-l_1)\omega_0+(k_2-l_2)\nu_0} \mbox{~in~}\chi_{r+1}
$$
Thus, the whole scheme of the computation of the Moser normal form
becomes a sequence of basically trivial algebraic operations, i.e.
multiplication or division by constants and computations of Poisson
brackets for polynomial functions. Let us note that the convergence
of the series is based on the fact that the method introduces
divisors of the form $im_1\omega_0+m_2\nu_0$, with $(m_1,m_2)$
integers, which can never become very small.\\
\\
\noindent{\it (iii) Normal form dynamics.} The final expression of
the Moser normal form is the Hamiltonian function
$Z(a,\xi,b,\eta)=H^{(N_t)}(a,\xi,b,\eta)$. This function has the
form:
\begin{eqnarray}\label{nfhyp}
Z(I=iab,c=\xi\eta)=i \omega_0 ab+\nu_0\xi\eta + \zeta_{21}
a^2b^2+\zeta_{22}\xi^2 \eta^2\nonumber \\
+\zeta_{23} a b \xi \eta+\zeta_{31} a^3b^3
+\zeta_{32} a b\xi^2\eta^2\nonumber \\
+\zeta_{33}q^2p^2 \xi\eta+\zeta_{34}\xi^3\eta^3+...
\end{eqnarray}
with terms depending on powers of the products $I=iab,c=\xi\eta$ up
to order $N_t/2+1$ (for symmetry reasons in the original Hamiltonian
$N_t$ has to be chosen even). By Hamilton's equations we trivially find
$\dot{I}=\dot{c}=0$. Thus, both quantities $(I,c)$ represent
integrals of motion under the normal form dynamics. In contrast to
what happens with the usual Birkhoff series (see Contopoulos 2002
for a review), we emphasize that the integrals $I,c$ in the above
computation are not only formal. In fact, the series giving them are
convergent. Thus, the integrals represent true invariants of motion,
whose precision of computation within the Moser domain of
convergence depends only on the level of the series truncation.
The physical meaning of these integrals is the following:

(a) The integral $I$ is given by $I=J+$higher order terms, where, as
noted above, $J$ is the action integral of the harmonic oscillator
in the elliptic degree of freedom of the linearized equations of
motion. Being produced by the full equations of motion, $I$ is the
action integral of a nonlinear oscillation, which, as explained in
section 2, represents an independent oscillation taking place in the
{\it center manifold} of the unstable equilibrium point $L_1$. If we
set $\xi=\eta=0$, one such oscillatory solution corresponds to one
member of the short-period orbit around the point $L_1$. Thus, $I$
is a label of the whole family of these orbits, with an increasing
value of $I$ representing an increasing size of the epicycle
described by the short-period orbit around $L_1$. In particular, the
value $I=0$ represents the limit when the size of the epicycle
reduces to zero, i.e., the Lagrangian point $L_1$ itself.

(b) The integral $c=\xi\eta$ yields a family of invariant
hyperbolae in the plane $(\xi,\eta)$. Consider a fixed value
of $I$. For every point $(\xi,\eta)$ within the Moser domain of
convergence, using the transformation equations (\ref{phitra})
we can find a corresponding point in the original variables.
Then, the points on one invariant hyperbola are mapped on
points on an invariant curve in the phase space of the
original variables. This curve is hereafter called a `Moser
invariant curve'. Such curves characterize the structure of
{\it chaotic orbits} in the vicinity of the unstable equilbrium.
In particular, all the chaotic orbits have their consequents
arranged along such curves (see \citet{b10,b20,b18}
for a detailed discussion of the properties of the Moser curves
in simple dynamical systems). Of particular importance is the
value $c=0$. Then, one Moser curve splits in three parts, namely
(1) the invariant point $\xi=\eta=0$, i.e., the fixed point of
a short-period orbit, (2) the $\xi-$axis ($\eta=0$), i.e., the
unstable manifold, and (3) the $\eta-$axis ($\xi=0$), i.e.,
the stable manifold of the short-period orbit. These are reduced
to the fixed point and stable and unstable manifolds of the
Lagrangian point $L_1$ itself for $I=0$.

After the above definitions, we discuss now our main result, namely
the connection between the Moser domain of convergence and the
chaotic spiral arms in our galactic models.

\subsection{Moser domain of convergence}

\begin{figure*}
\centering
\includegraphics[scale=0.29]{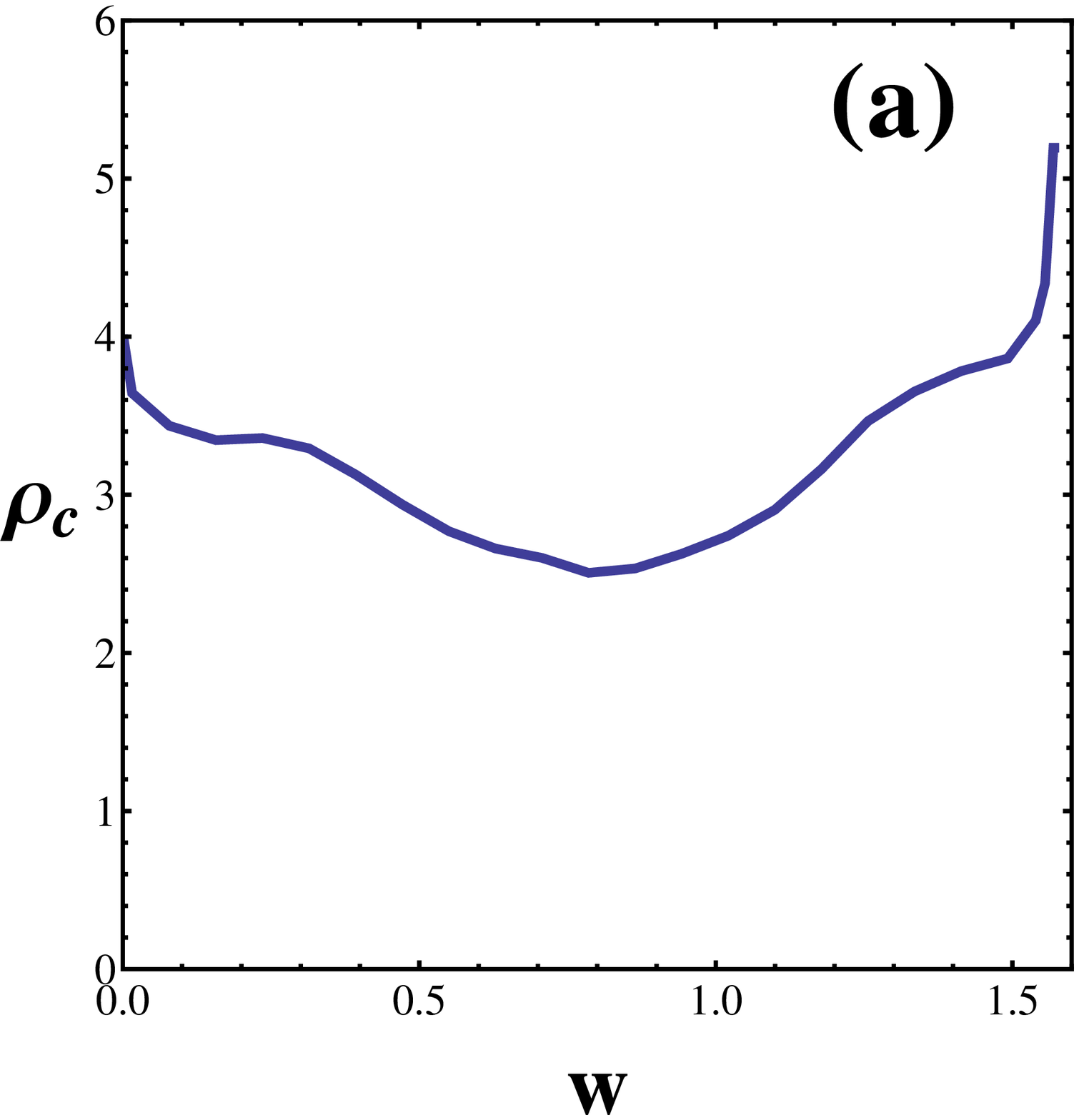}
\includegraphics[scale=0.29]{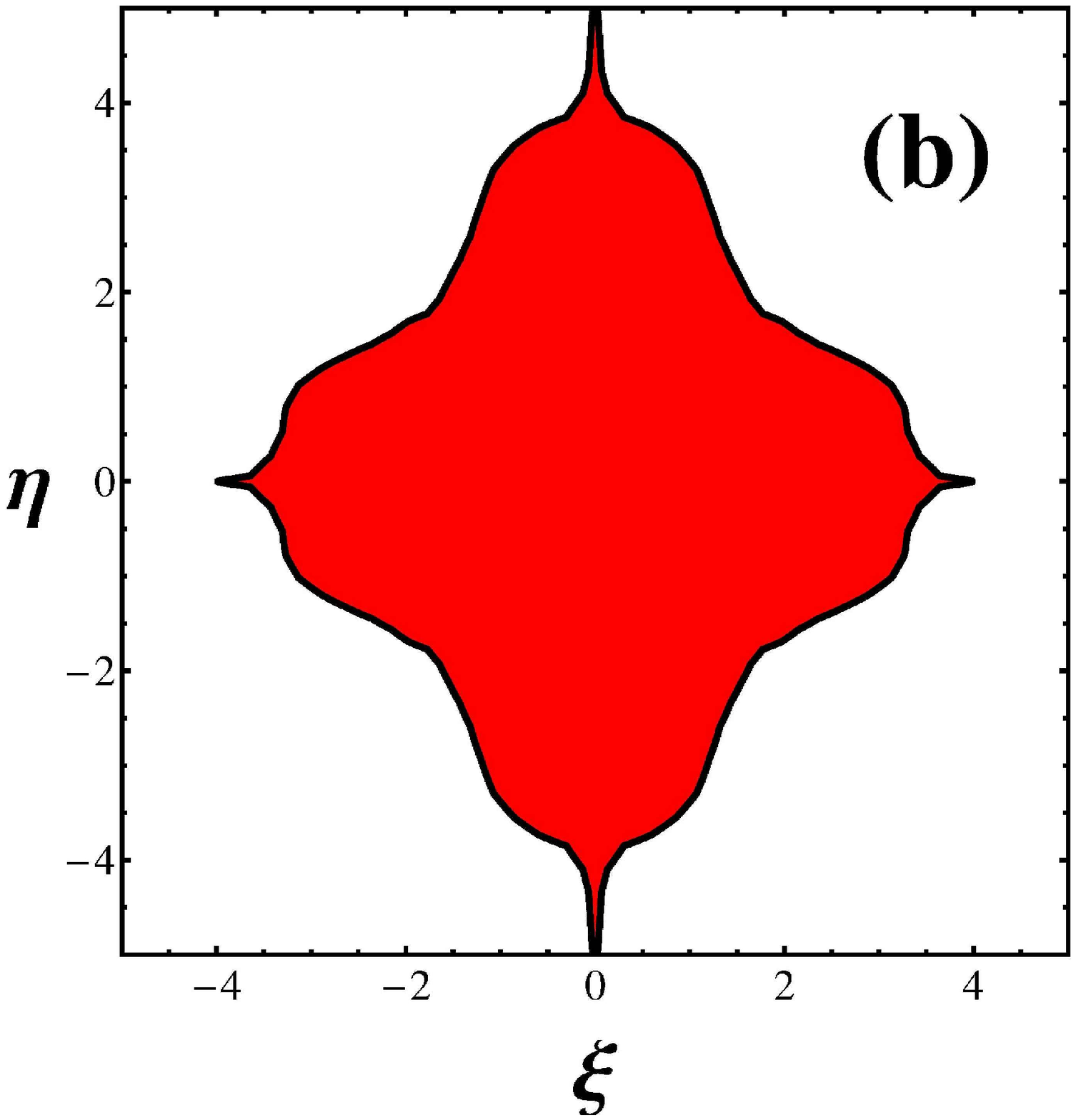}
\caption{(a) The radius of convergence of the series (\ref{phitra})
as a function of the angle $w$ for $w=0$ to $w=\pi/2$.(b) The region
of convergence on the ($\xi,\eta$) plane. Both (a) and (b)
correspond to the galactic model "C".} \label{radconv}
\end{figure*}

In the sequel, we focus on the case $I=0$, i.e., the computation of
the Moser domain of convergence around the Lagrangian point $L_1$
itself. However, the same method can be applied to any case with
$I\neq 0$.

Setting $a=b=0$ in the transformation series (\ref{phitra}),
all four series become polynomial in the two variables $(\xi,\eta)$.
In order to compute the domain of convergence we now work with a
variant of the method introduced in a previous paper \citep{b18}.
From Eq.(\ref{phitra}), taking, as an example, the truncated series
$\Phi_u(a=0,b=0,\xi,\eta)$, we have
\begin{equation}\label{phiuser}
\Phi_u(a=0,b=0,\xi,\eta)=\sum_{s=1}^{N_t+2}\sum_{k=0}^s
f_{s,k}\xi^k\eta^{s-k}
\end{equation}
Define now a particular direction in the $(\xi,\eta)$ plane
passing through the origin, parameterized by the equations
$\xi=\rho\cos w$, $\eta=\rho\sin w$, with $w$ fixed. Then, the
integral $c=\xi~\eta$ (see previous subsection) becomes
$c=\rho^2\cos w\sin w$. Substituting these expressions in the
series (\ref{phiuser}), we obtain a series depending only on
powers of the distance $\rho$ from the origin:
\begin{equation}\label{phiuser2}
\Phi_{u}(\rho;w)=\sum_{s=1}^{N_t+2}B_s(w)\rho^s
\end{equation}
with
\begin{equation}\label{bepol}
B_s(w)=\sum_{k=0}^sf_{s,k}\cos^k(w)\sin^{s-k}(w)
\end{equation}
We define the sequence of `Cauchy radii', depending on the integer
$s=1,2,...$ by
\begin{equation}\label{caur}
\rho_{\textrm{c,s}}(w) =|1/B_s(w)|^{1/s}
\end{equation}
According to the Cauchy theorem, the series (\ref{phiuser2})
converges inside the radius
$\rho_\textrm{c}(w)=\lim_{s\rightarrow\infty} \rho_{c,s}(w)$. In
practice, since we work with a finite truncation, we compute the
sequence (\ref{caur}) and check numerically that it converges to a
nearly constant value for large $s$. Introducing also a 0.95 safety
factor, we define a numerical estimate of the convergence radius as
\begin{equation}\label{caurnum}
\rho_{\textrm{c,num}}(w)=0.95 \rho_{c,N_t+2}(w)
\end{equation}
with $N_t=20$ in all computations below.

Figure \ref{radconv}a shows an example, for the galactic model
$"C"$, of the numerically computed radius of convergence
$\rho_{\textrm{c,num}}$ as a function of the angle $w$ where $w\in
[0,\pi/2]$. Figure \ref{radconv}b shows the domain of convergence
(red) inside the limiting black curve corresponding to the radius of
convergence for all four quadrants in the same example. Let us point
out that any one of the four series of Eq.(\ref{phitra}) can be used
in the computation of the convergence domain, since, by the series
construction, all transformations should converge within the same
domain.

\begin{figure*}
\centering
\includegraphics[scale=0.40]{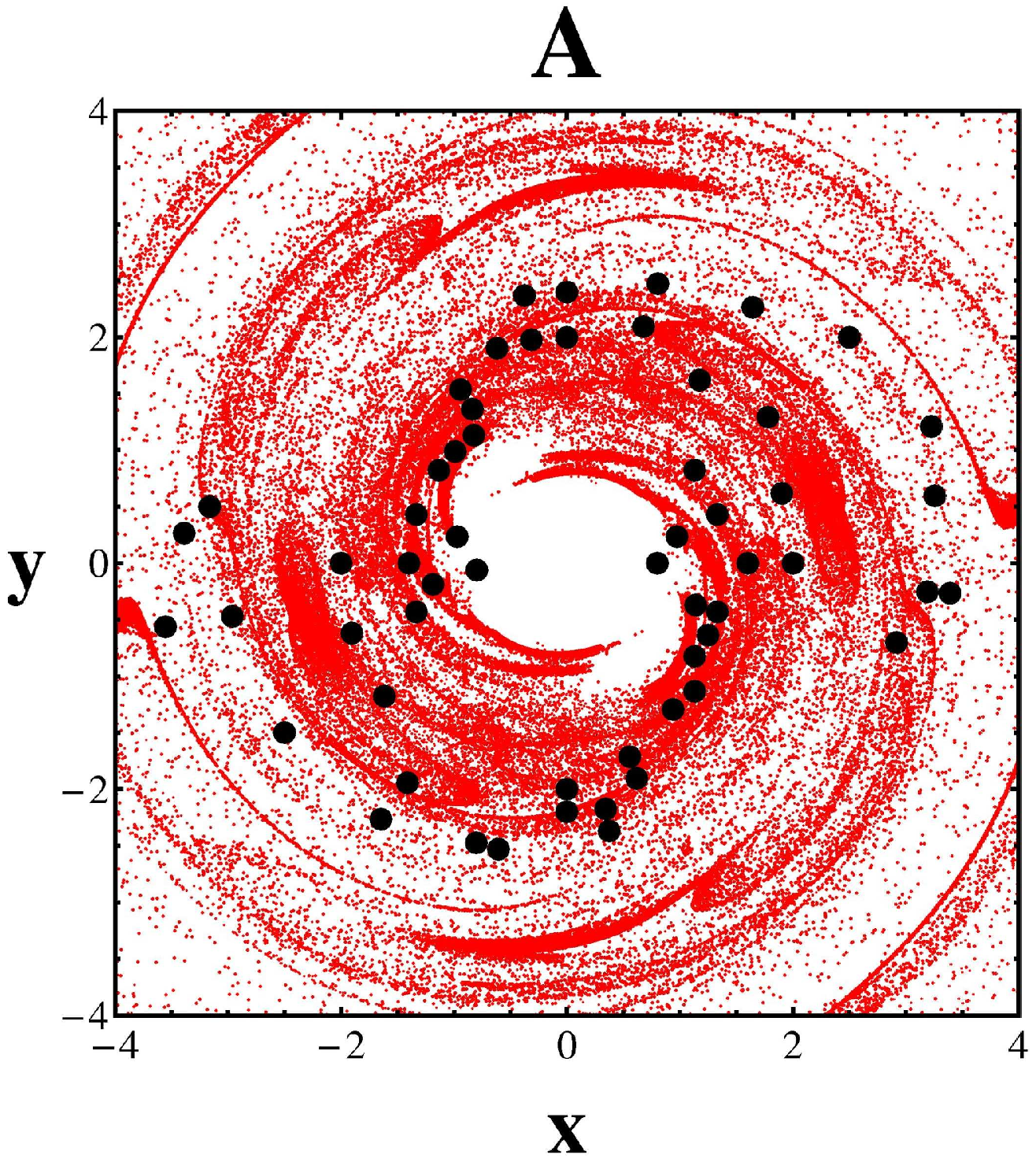}
\includegraphics[scale=0.40]{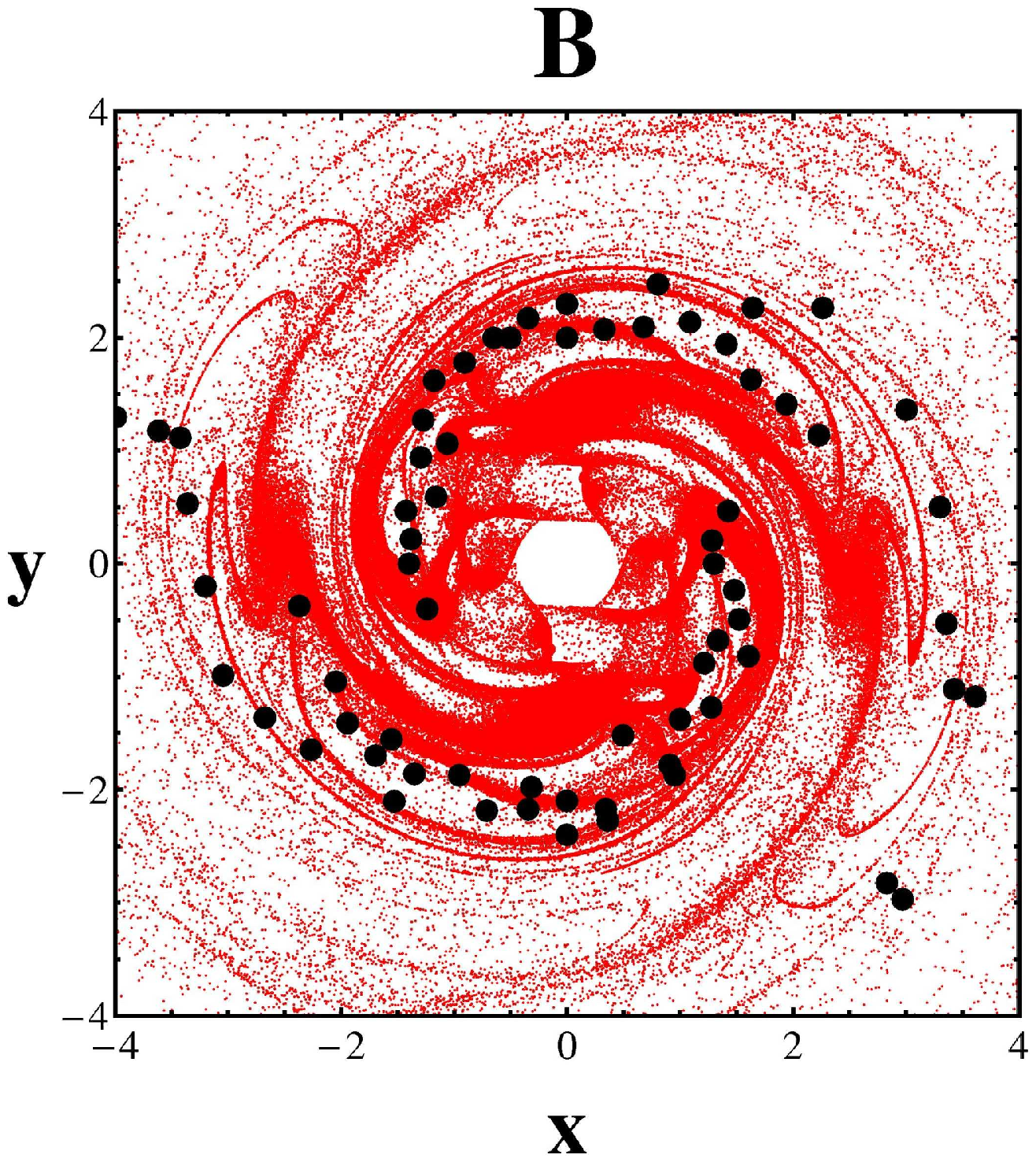}
\includegraphics[scale=0.40]{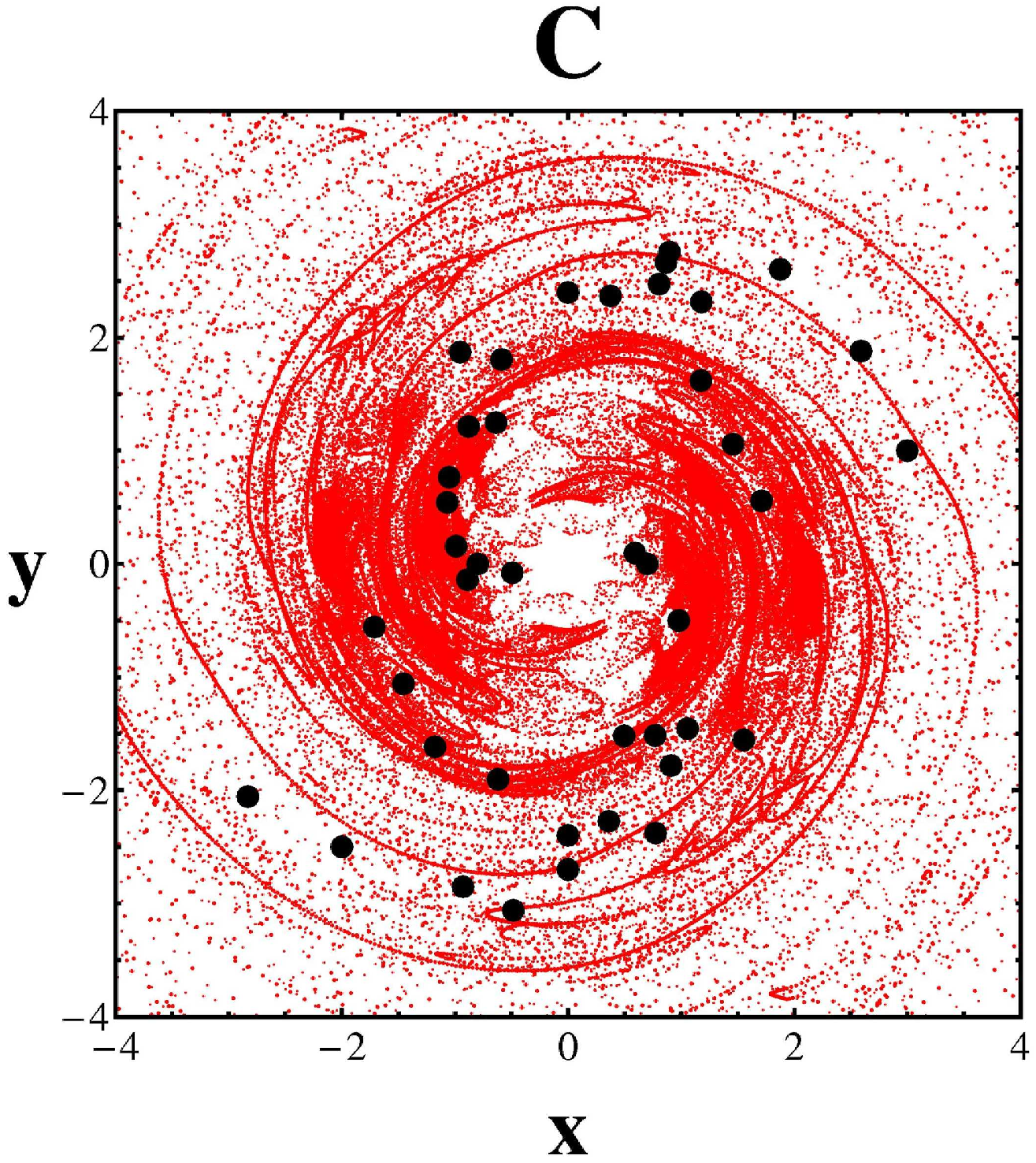}
\caption{The extended Moser domains of convergence ${\cal
M}_{ext}$ (see text), for our three galactic models $"A"$, $"B"$
and $"C"$, in the configuration space of the galaxy. We plot the
first three apocenters of each orbit with random initial condition
inside the Moser domain of convergence ${\cal M}$ (like the one of
Fig. \ref{radconv}b). One can identify spiral arms, in all three
galactic models, which are consistent with the local maxima of the
projected surface density of the N-body particles (black dots).
The dots are the same as in Fig.\ref{bqr2}.} \label{apomos}
\end{figure*}

\begin{figure*}
\centering
\includegraphics[scale=0.53]{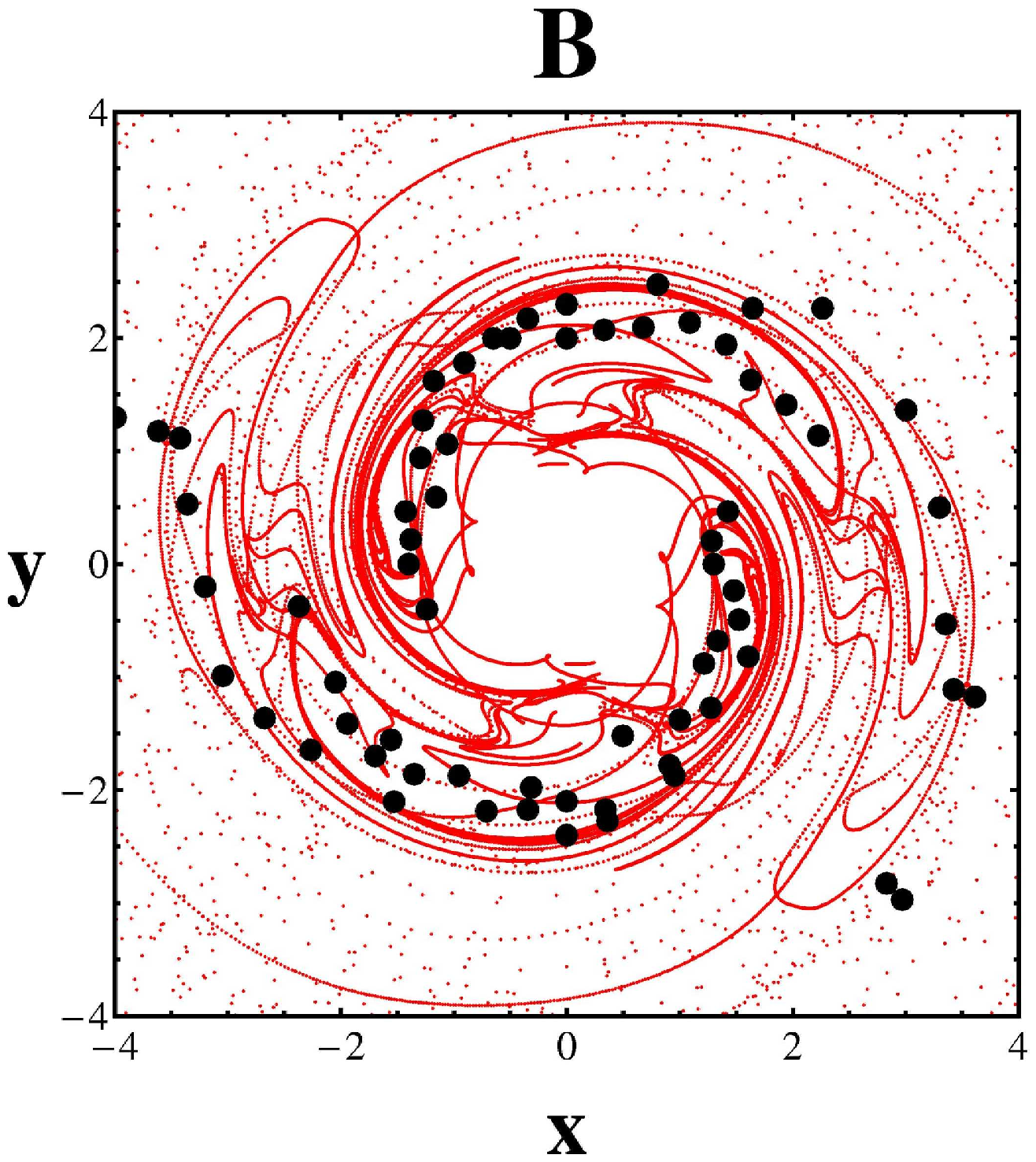}
\includegraphics[scale=0.40]{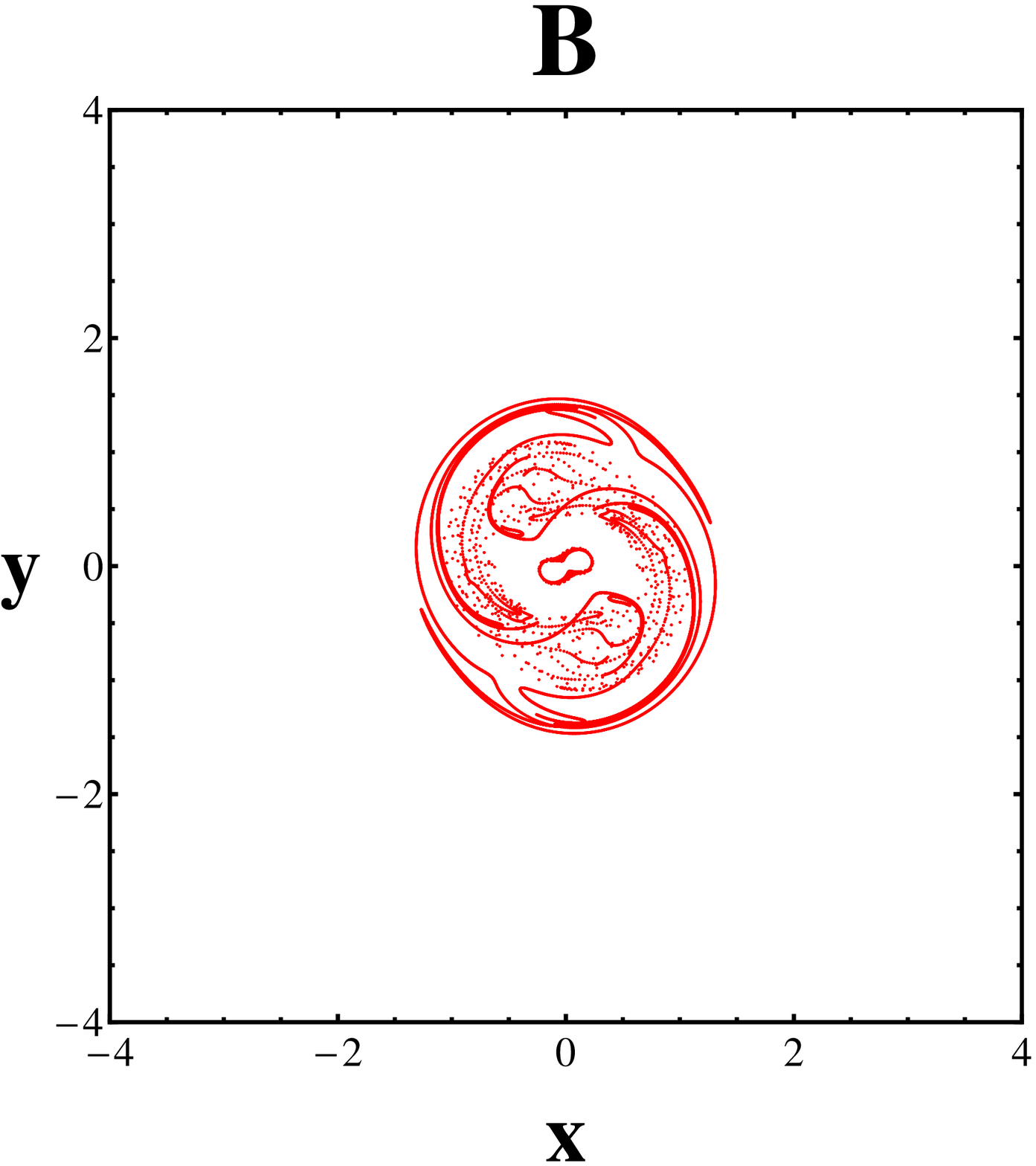}
\caption{\textbf{Left:} We plot the first three images on the
apocentric surface of section, of the initial conditions along the
$\xi$-axis of the Moser domain of convergence (which corresponds
to the unstable asymptotic curve of the $L_1$ (and its symmetric
$L_2$) point together with the local maxima of the projected
surface density of the N-body particles (black dots).
\textbf{Right}: same as in the left panel but for the pericentric
surface of section ($\dot{r}=P_r=0$, $\dot{p}_r>0$). }
\label{apopl1}
\end{figure*}
Note that the present case is somewhat different from the cases
considered previously in \citet{b20} and \citet{b18}. Namely, in
those studies it was found numerically that the limiting value of
$c=c_{\textrm{lim}}$ (see (iii) of subsection 3.1 for the
definition of $c$), was independent of the angle $w$. Here,
instead, we find that $c$ depends on $w$ so that the boundary of
the Moser domain differs from a pure hyperbola. In particular, we
find that $\rho_\textrm{c}$ is finite in both axes $\xi=0$ and
$\eta=0$. The origin of this difference is due to a difference
between the convergence domains in the case of real analytic
mappings on the plane, and systems, like the present one, produced
by a continuous Hamiltonian flow (see \citet{b10}).

Figure \ref{apomos} shows now the main result. It is produced as
follows: We construct a set of randomly distributed initial
conditions inside the Moser domain of convergence in the
($\xi,\eta$) plane (as in Fig.\ref{radconv}b). Using the
transformation equations (\ref{phitra}), with $a=b=0$, as well as
the linear transformation $\overline{A}$ of section 2, every one
of these points can be mapped to a point in the original canonical
variables $(r,\phi,P_r,P_\phi)$ and eventually in the cartesian
variables ($x,y,p_x,p_y$). This process defines the image of the
Moser domain in the phase space of the original variables. This is
hereafter called ${\cal M}$.

Every initial condition in $\cal {M}$, when integrated forward in
time, reaches consecutive apocentric positions of the orbit, at
some times $t=t_1,t_2,...$. Hereafter, we call apocentric surface
of section the surface defined by the relation $\dot{r}=P_r=0$,
$\dot{p}_r<0$. We note that this is a 2D surface of section
embedded in the 4D phase space. In subsequent plots we focus on
the projection of this surface in the usual configuration space
($x=r \cos(\phi)$, $y=r \sin(\phi)$).

Following, now, \citet{b8} (see also \citet{b10}), we hereafter
call {\it the extended Moser domain} in the plane $(\xi,\eta)$ the
union of the original Moser domain of Fig.\ref{radconv}b along
with all its {\it forward} images at the times $t_1,t_2,...$. The
latter can be computed analytically, given the normal form of
Eq.(\ref{nfhyp}). This process allows to establish an extension of
the transformations of Eq.(\ref{phitra}) from the plane
$(\xi,\eta)$ to the configuration space $(x,y)$. The extended
Moser domain is found by propagating forward in time all the grid
points in ${\cal M}$. The propagation can be done analytically
using a so-called ``extended method'' developed in \citet{b10},
but here, for simplicity, we simply perform it by numerical
integration of the orbits.

We hereafter denote by ${\cal M}_{ext}$ the image of the extended
Moser domain on the plane $(x,y)$. Figure \ref{apomos} shows the
intersection of ${\cal M}_{ext}$ with the apocentric surface of
section $\dot{r}=0$, $\dot{p}_r<0$, with a computation of ${\cal
M}_{ext}$ up to a time covering three apocentric passages for all
the orbits in ${\cal M}$. The reason for this choice of the
apocentric section is that ${\cal M}_{ext}$ in this section
contains the apocentric sections of the unstable manifolds of
$L_1$ and their neighborhoods.

As shown clearly in Fig.\ref{apomos}, the images of the Moser
domains of convergence in the configuration space of all three
models define areas on non-zero measure which have the forms of
spiral arms,  which are consistent with the images of the local
maxima of the projected surface densities of the N-body particles
(black dots). In fact, a careful inspection of all three panels in
Fig.\ref{apomos} reveals that there are domains where distinct
parts of ${\cal M}_{ext}$ overlap. This is allowed since the
transformation (\ref{phitra}) is not bijective. Actually, the
greatest enhancement of the spiral densities occurs, precisely, in
domains of such overlapping.

It is emphasized that both the N-body particles forming the spiral
arms as well as fictitious particles with initial conditions in the
set ${\cal M}_{ext}$ move along chaotic orbits. A theoretical
interpretation of the role of ${\cal M}_{ext}$ in determining the
dynamics along the chaotic spirals is given in the next section.

In Fig.\ref{apopl1} we plot the images of initial conditions along
the $\xi$-axis of the Moser domain of convergence, for the
galactic model "B".
 The images of
these initial conditions correspond to the unstable asymptotic
curves of the equilibrium points $L_1$, $L_2$. In the left panel
of figure \ref{apopl1} we plot the first three apocenters (red
dots) of the unstable asymptotic curves, superposed to the local
density maxima of the N-body particles (thick black dots). Note
that, besides the main spiral structure, these plots indicate that
the manifolds support also a second pair of spiral arms nearly
parallel to the main spiral arms. A similar result was found in a
previous paper using orbital structure study (see Fig. 20 of
Contopoulos and Harsoula 2013.) Let us note that the "double
spiral" structure, is a notable morphological feature in many
barred-spiral galaxies.

In the right panel of figure \ref{apopl1} we plot the pericenters
of the unstable asymptotic curves from $L_1$ and $L_2$, which
support the limit of the bar and the innermost part of the spiral
structure.

\section{Theoretical interpretation}

A typical property of all galactic dynamical systems with a strong
bar is that the phase space beyond corotation is open to escapes.
Numerical simulations show that most stars in chaotic orbits
acquire escape velocities from the galaxy in rather short
timescales (of the order of a few dynamical periods only). On the
other hand, the stars with initial conditions close to the
phase-space invariant structures such as invariant manifolds or
cantori are "sticky", i.e they resist in general the escaping flow
for longer times, which are often sufficient to support structures
such as chaotic spiral arms.

Our interpretation of the role that the Moser domains play in the
phenomenon of chaotic spirals is based on the findings in the recent
work of \citet{b18}. In this work, the following two properties were
demonstrated:

i) All the chaotic orbits with initial conditions inside a Moser
domain of convergence remain bounded within the extended Moser
domain for arbitrarily long times. This property is a consequence of
the Moser normal form dynamics. Namely, the successive consequents
of the chaotic orbits with initial conditions within the Moser
domain necessarily lie in one invariant Moser curve (i.e. a
hyperbola $\xi\eta=c$ in the $(\xi,\eta)$ plane and its image in the
configuration plane).

ii) The boundary of the Moser domain acts as an attractor for all
the chaotic orbits with initial conditions outside but close to it,
although these orbits escape asymptotically to infinity. This
property implies that the chaotic escapes do not take place in
random directions in phase space, but the successive consequents of
the escaping orbits necessarily approach closer and closer to the
boundaries of one or more Moser domains (formed around one or more
unstable periodic orbits in the same system, see \citet{b18}). As a
result, the preferential directions of escape for all the orbits are
those along which the Moser domains of the unstable periodic orbits
extend  to infinity.

\begin{figure*}
\centering
\includegraphics[scale=0.53]{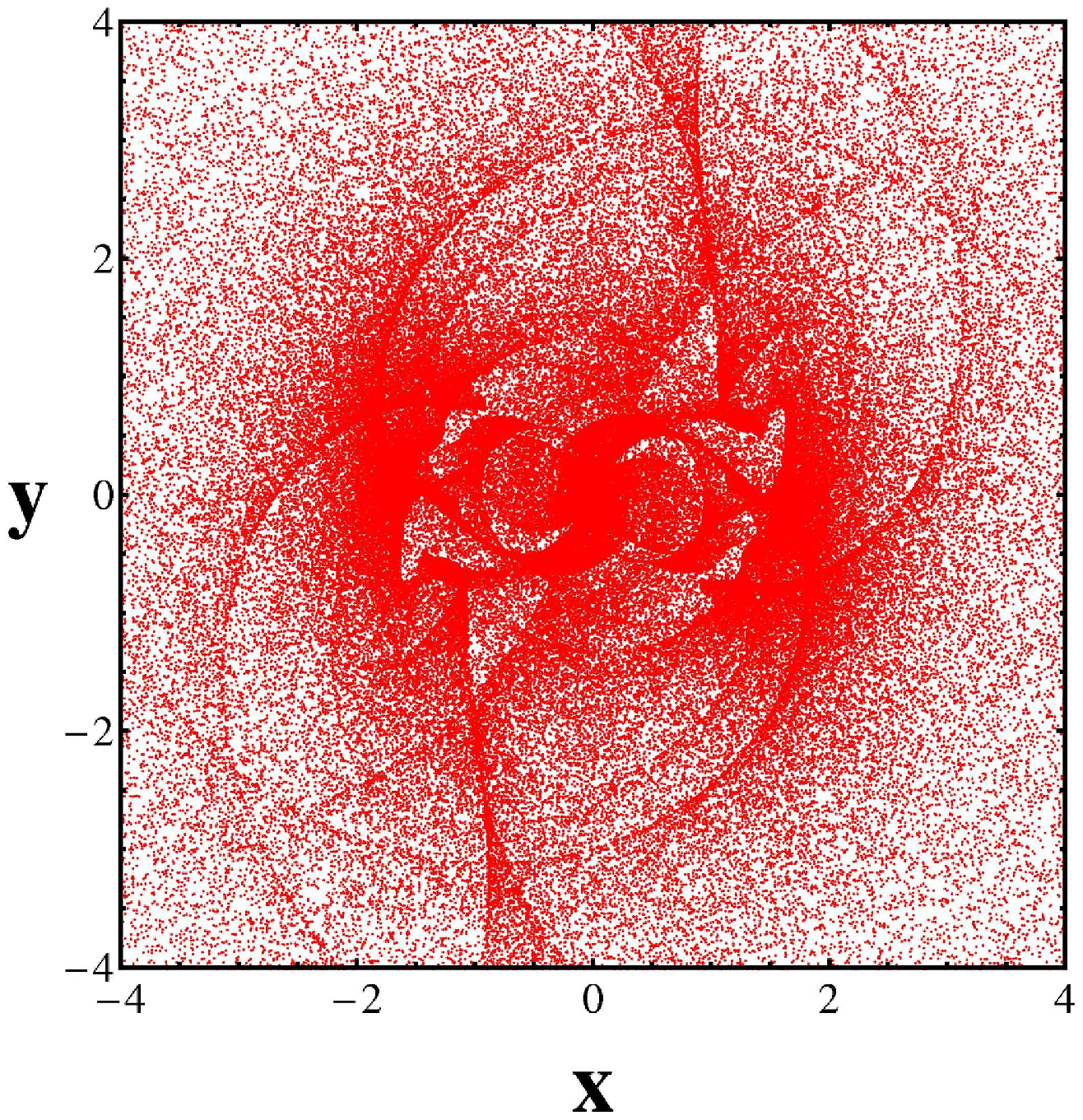}
\includegraphics[scale=0.40]{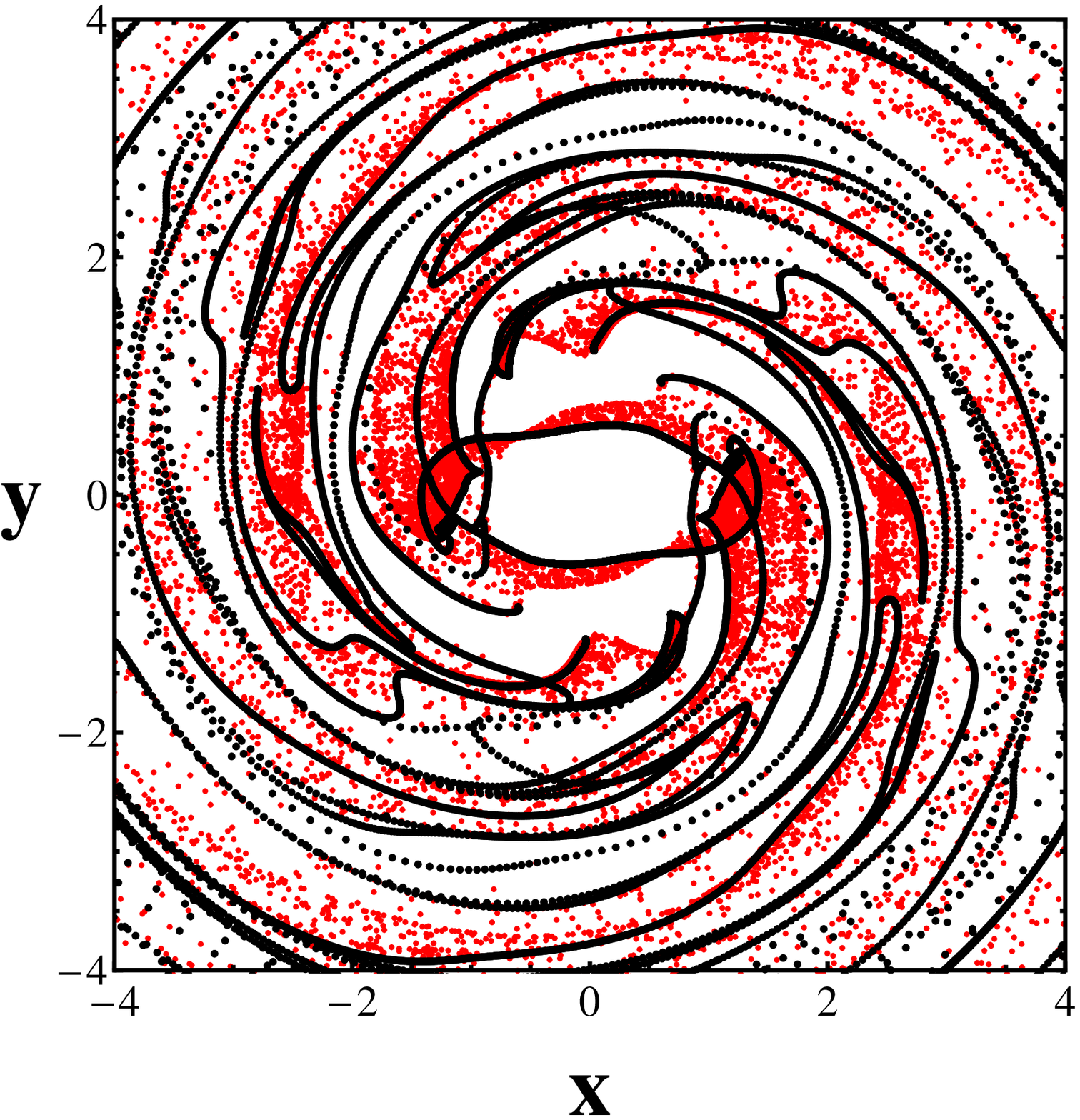}
\caption{ \textbf{Left}: The projection on the configuration plane
of the galactic model "B" of orbits whose initial conditions on
the $(\xi,\eta)$ plane, are inside the grid
($-5<\xi<5$,$-5<\eta<5$), but outside the Moser region of
convergence (red region of Fig. \ref{radconv}). \textbf{Right}:
The first apocentric positions of the orbits of (a) together with
the first apocenter of the image of the boundary of the Moser
domain of convergence of Fig. \ref{radconv}  (black curve). These
chaotic orbits will finally escape with a slow diffusion along
spirals.} \label{mosesc}
\end{figure*}

The above results were found in simple area-preserving mappings,
but we now show how they translate in the case of the Moser
domains ${\cal M}_{ext}$ computed in our galactic models. Figure
\ref{mosesc} summarizes the relevant information. The left panel
of Fig. \ref{mosesc} gives the projection on the configuration
space of the galactic model "B", of orbits whose initial
conditions on the $(\xi,\eta)$ plane are inside a grid
($-5<\xi<5$,$-5<\eta<5$), but outside the Moser domain of
convergence (red region of Fig. \ref{radconv}). Using the
transformation equations (\ref{phitra}), with $a=b=0$, as well as
the linear transformation $\overline{A}$ of section 2, we map
these points in the original canonical variables
$(r,\phi,P_r,P_\phi)$ and eventually in the cartesian variables
($x,y,p_x,p_y$). Using these initial conditions, we then integrate
the orbits until they reach their first apocentric section (we
only consider the orbits which have initially a negative energy in
the inertial frame, i.e.
$E=0.5(P^2_r+P^2_{\phi}/r^2)+V(r,\phi)<0$, since the orbits with
$E>0$ escape from the system immediately).  The so-resulting
distribution of the orbits in the apocentric section is shown in
the right panel of Fig. \ref{mosesc}, together with the image , in
the same section, of the boundary of the Moser domain (black
curve).

   We observe that the boundary of the Moser
domain ${\cal M}_{ext}$ attracts all the exterior orbits in its
neighborhood. These orbits follow escaping paths close to this
boundary, along the spiral pattern. In fact, this spiral makes
several revolutions as shown in Fig.\ref{mosesc}b. However the
density of points falls (nearly exponentially) as the distance
from the center increases, thus practically limiting the extent
along which the spiral arms are traced by an appreciable amount of
matter.

On the other hand, we may note that a clear, albeit only
qualitatively correct, theoretical picture can be obtained by
constructing an approximate explicit {\it mapping} model to
represent the dynamics in the corotation region around the
Lagrangian points. We close our analysis in this paper by showing
results based on such an approximate mapping, which we constructed
using a method borrowed from solar system studies (the so-called
`Hadjidemetriou method' \citep{b40,b19}). Deferring all technical
details of the mapping construction to Appendix B, we here summarize
only the final result. By constructing a so-called ``averaged
Hamiltonian'' based on the epicyclic approximation applied to the
potential of each N-body galactic model, we end up by showing that
the dynamics around the Lagrangian points $L_1$ and $L_2$ can be
approximated by a version of the well known Chirikov standard map
\citep{b46}:
\begin{eqnarray}\label{stand}
~~~~~~~~~~~~~\Theta'=~\Theta + Y'\nonumber
\\~~~~~~~~~~~~~Y'=Y +K \sin(\Theta)
\end{eqnarray}
where $K$ is a non-linearity parameter depending on the perturbation
of each galactic model. The variables $(\Theta,Y)$ in the mapping
(\ref{stand}) are connected to cylindrical coordinates via the
azimuth $\phi$ and its conjugate action $J_{\phi}$, which measures
the distance away from corotation ($J_\phi>0$ outside corotation and
$J_{\phi}<0$ inside corotation). The value of the non-linearity
parameter $K$ is proportional to the strength of the $m=2$ component
of the potential at corotation. The value of $K$ derived for the
three different models considered is $K_A=2.7$ (model $"A"$),
$K_B=4.6$ (model $"B"$) and $K_C=9.3$ (model $"C"$) (see Appendix B
for details). Thus, in all three models the non-linearity is quite
strong, and results in a phase space where most chaotic orbits are
free to escape.

\begin{figure*}
\centering
\includegraphics[scale=0.68]{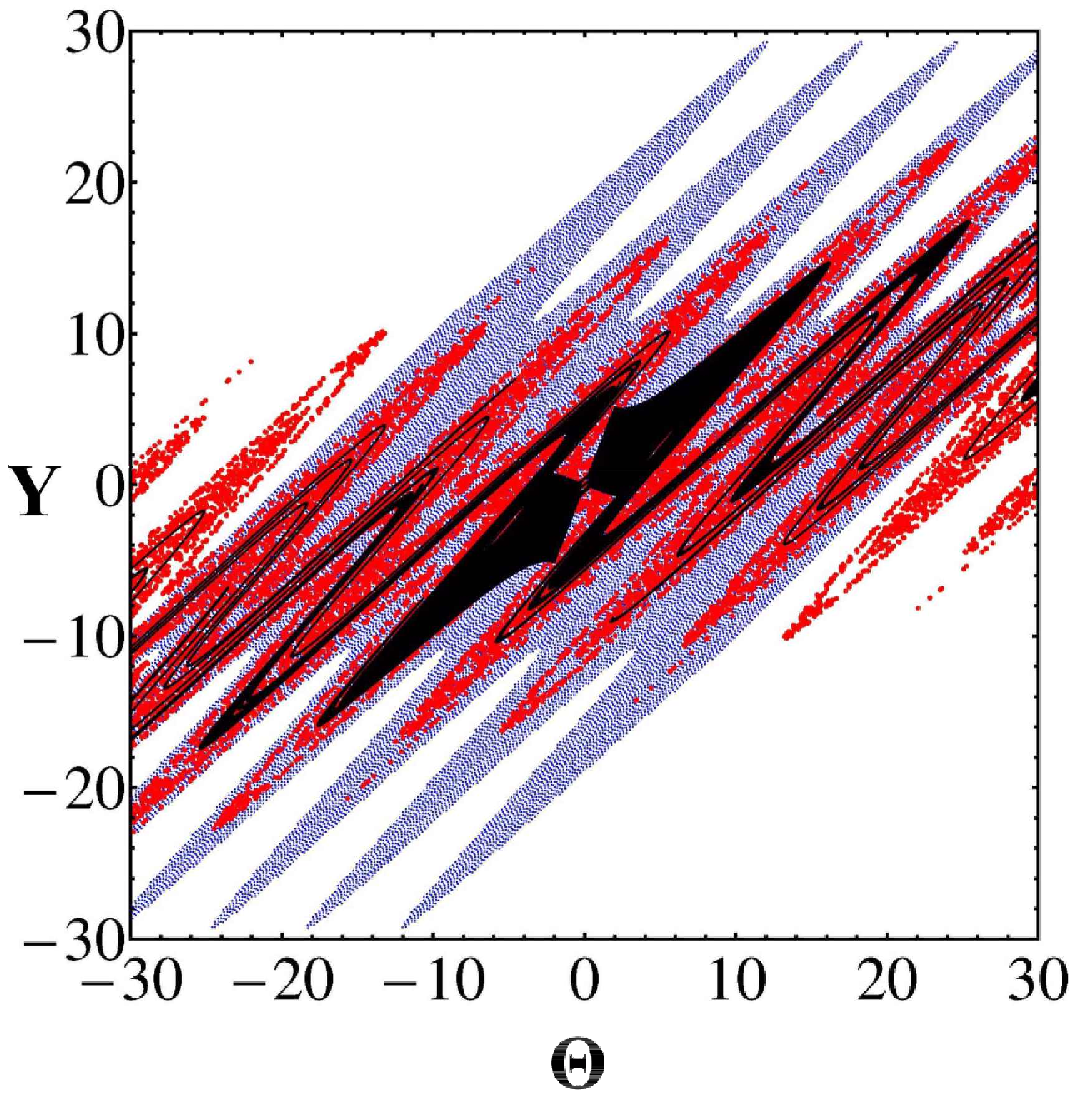}
\includegraphics[scale=0.40]{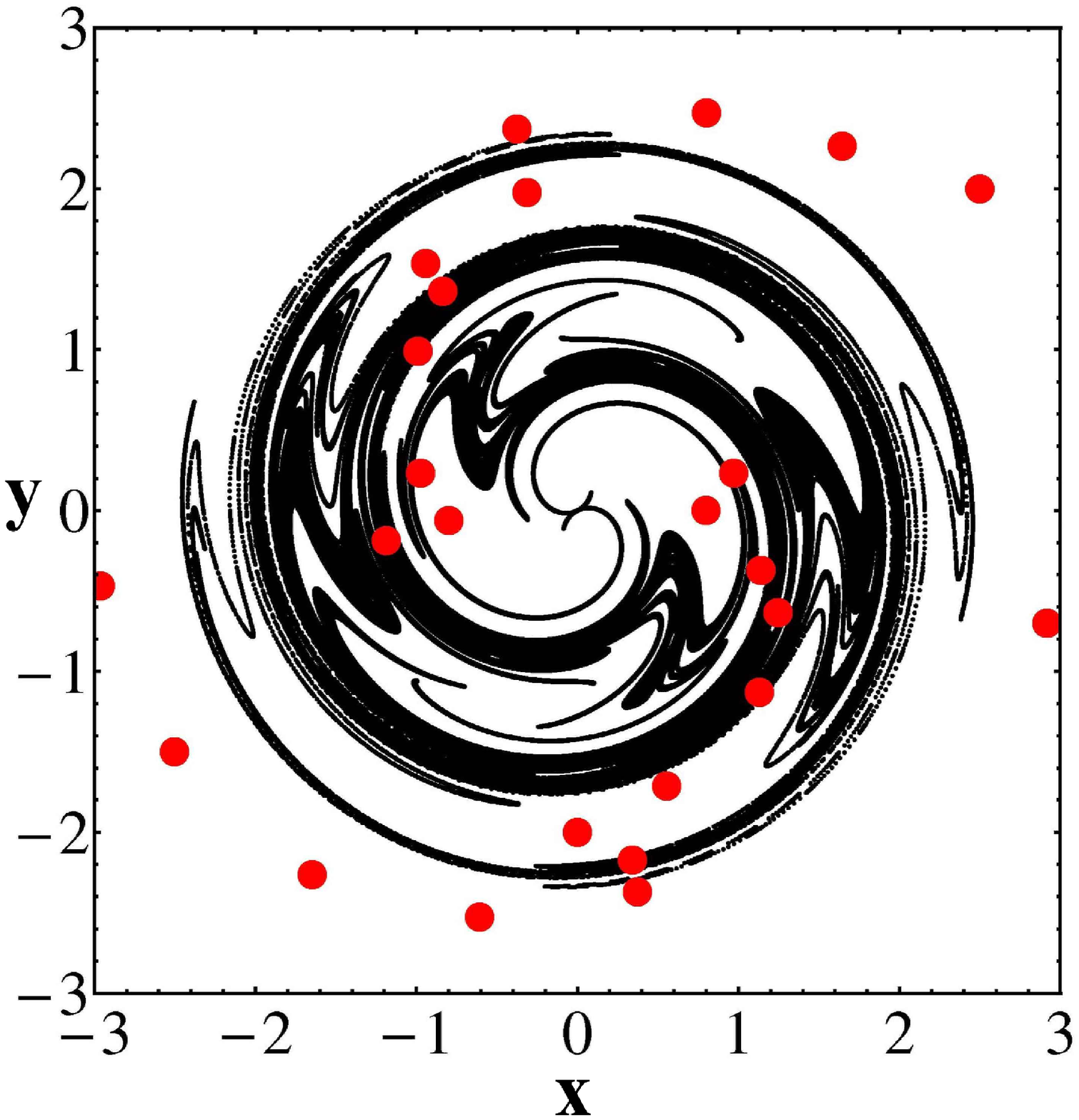}
\caption{(a) The successive iterations of a grid of points outside
the region of convergence (black region) of the standard map
(\ref{stand}) for the galactic model "A". Blue is the first and red
is the third iteration of this region. It is obvious that the
successive iterations get closer and closer to the black region of
convergence. In fact, the outer boundary of the Moser region of
convergence acts like an attractor in the phase space. (b) The Moser
domain of convergence (black region of (a)) in the configuration
space of the galactic model $A$, superimposed with the local maxima
of the projected surface density (red dots). It's obvious that the
spiral arms of the mapping approximation are more tightly wounded
than the ones of the galactic model.} \label{mossta}
\end{figure*}

The Lagrangian points $L_1$, $L_2$, or the fixed points of the
family of the short-period orbits $PL_1$, $PL_2$, correspond to the
hyperbolic point $(\Theta=0,Y=0)$ (or $(2\pi,0)$ which is the same
point modulo $2\pi$). Using the same formulas for the production of
the Moser normal form for area-symplectic mappings as in
\citep{b18}, we compute the Moser domain of convergence of the
mapping (\ref{stand}) first in the mapping variables $(\Theta,Y)$,
and then in the original cylindrical canonical variables of our
galactic models.

Figure \ref{mossta}a shows an example of the Moser domain of
convergence (black) in the mapping variables $(\Theta,Y)$, for the
galactic model $"A"$. Also, taking a set of initial conditions
outside the Moser domain, the same plot shows their first (blue) and
third (red) iterations in the same plane. It is obvious that the
successive mappings of the initial conditions outside the Moser
domain of convergence come closer and closer to the boundary of the
(black) domain of convergence. However, these orbits can only
approach asymptotically the boundary of the domain, and they cannot
enter inside the the black domain. Hence, the boundary of the Moser
domain acts like an attractor in the phase space for all the orbits
with initial conditions outside the Moser domain. These orbits
finally escape to infinity. In fact the boundary of the Moser domain
(black) forms an infinity of oscillations beyond the limits of
Fig.\ref{mossta}a. Thus the orbits outside and close this boundary
extend to arbitrarily large ($\Theta$) and the corresponding spirals
of Fig. \ref{mossta}b extend to arbitrarily large distances.

Finally, Fig.\ref{mossta}b shows the image of Fig.\ref{mossta}a in
the configuration space of the galactic model $"A"$. Thus,
Fig.\ref{mossta} gives the same results as Fig.\ref{mosesc}, but
depicts more clearly the attraction of the escaping orbits by the
boundary of the Moser domain. Note that also in this simple mapping
model, the Moser domain exhibits a spiral form. However, we stress
that the mapping model (\ref{stand}) only serves for a theoretical
interpretation of previous results, while its comparison with the
exact model can only be qualitative. In fact, the spiral structure
in Fig.\ref{mossta}b appears more tightly wound than the true spiral
structure of the model (red dots).

Similar results were found also in the models $"B"$ and $"C"$.

\section{Conclusions}

In the present paper we demonstrate a close connection between the
form of the chaotic spiral arms in barred galaxies and an analytical
theory describing the chaotic orbits in the neighborhood of the
unstable points $L_1$ and $L_2$ at the end of bars due to
\citet{b2,b3}. Our present results complement in an essential way
the manifold theory of spiral structure \citep{b34,b27}, and they
allow to build analytically domains of non-zero measure in phase
space which correspond to a non-zero phase space density of stars
moving along the spiral arms. In particular:

1) We computed the so-called 'Moser normal form' (see section 3),
i.e. a {\it convergent} series of perturbation theory allowing to
characterize analytically the {\it chaotic} orbits with initial
conditions in the neighborhood of the invariant manifolds of the
unstable points $L_1$ or $L_2$. We gave particular examples of Moser
normal form computations based on potential models derived by N-body
simulations of barred-spiral galaxies.

2) We computed the domain of convergence of the Moser series in
the normal form variables, and found its image in the usual
configuration space of the disc plane. This image, in all three
galactic models has the form of trailing spiral arms, which can be
computed analytically knowing only the coefficients of the
potential expansion. We emphasized that when the orbits within a
Moser domain are recurrent, the spiral structure is formed by the
intersection of the domain of convergence with a so-called
`apocentric' section (see section 3).

3) We computed the local maxima of the surface density on the disc
for the real N-body particles and verified their good agreement with
the analytically computed spirals.

4) We gave a theoretical interpretation of this agreement (section
4) based on findings in previous works \citep{b20,b18} regarding the
dynamical role of the Moser domains in the stickiness and escape
dynamics in simple mappings with a phase space open to escaping
chaotic motions. We demonstrate that the boundaries of the Moser
domains of convergence act as attractors for the escaping chaotic
orbits with initial conditions near, but in the exterior of this
domain. On the other hand, all the chaotic orbits with initial
conditions inside a Moser domain necessarily reproduce the spiral
form of this domain, since they can never escape outside this
domain.

5) Finally, we constructed a simple mapping of the type of
Chirikov's standard map, based on the averaged-Hamiltonian approach
of \citet{b40,b19}, which allows to reproduce qualitatively the
apocentric section dynamics of the chaotic orbits in the
neighborhood of $L_1$ or $L_2$. The Moser domain of convergence
extends to infinity along the invariant manifolds of the mapping's
unstable fixed point at the origin. Thus, the geometric loci of the
corresponing spiral arms in the galactic plane extend to infinity.
However, the density of matter falls exponentially along these loci,
hence, the Moser theory leads to theoretical spiral arms of only a
finite extent beyond the corotation region.


\section*{Acknowledgements} We acknowledge support by the research
committee of the Academy of Athens through the project 200/854.

\newpage
\section*{Appendix A: The coefficients of the galactic potentials}
\begin{figure*}
\centering
\includegraphics[scale=0.20]{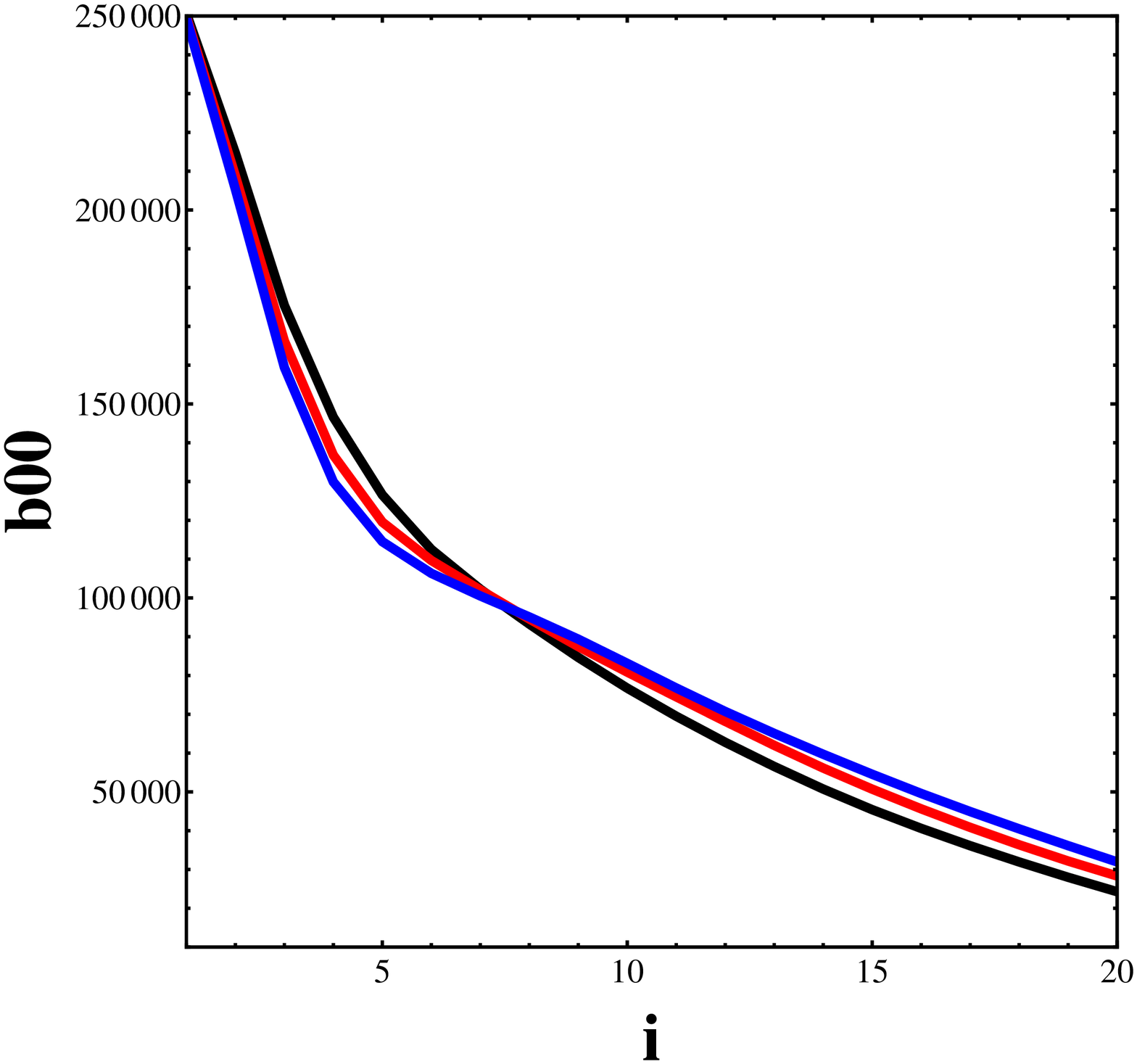}
\includegraphics[scale=0.20]{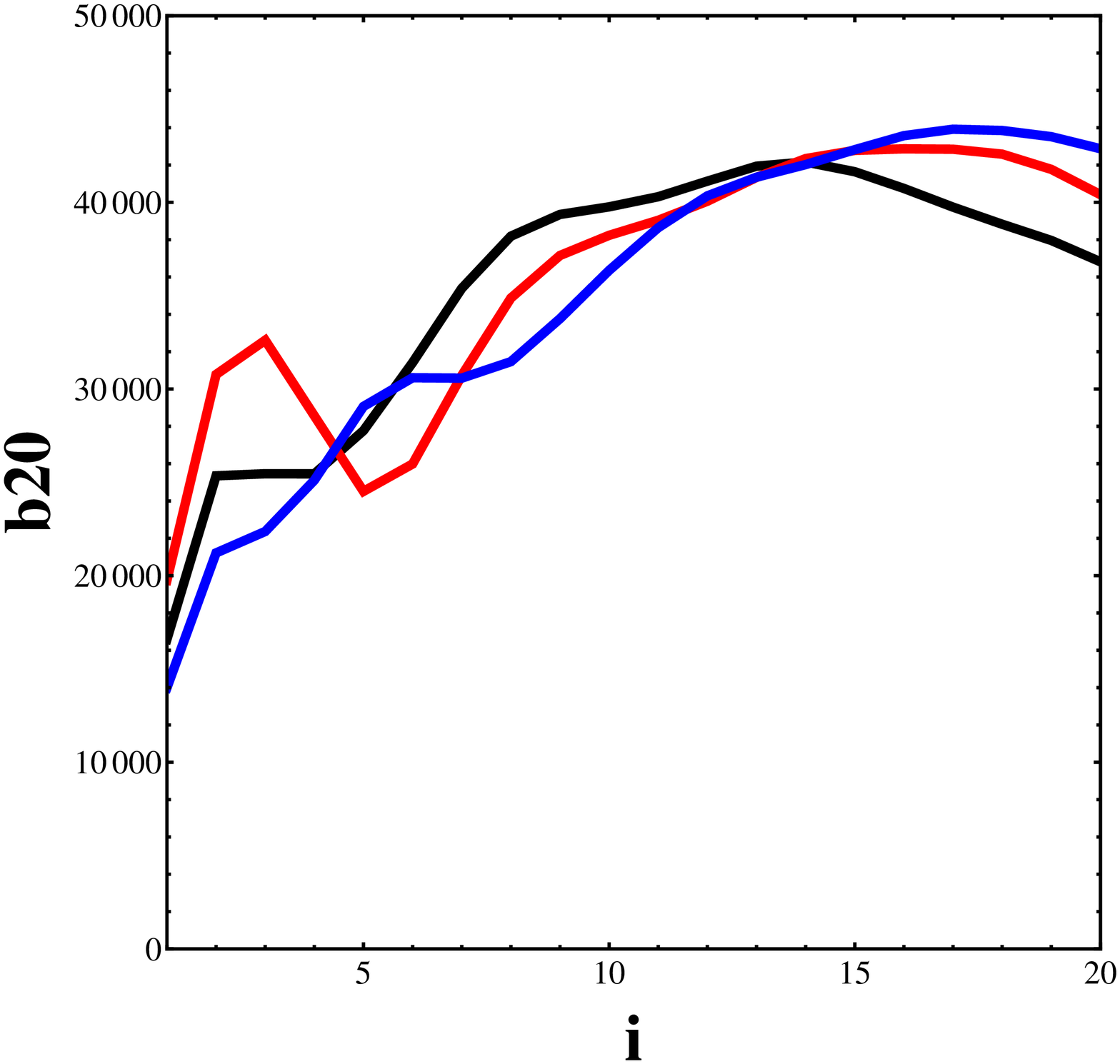}
\includegraphics[scale=0.20]{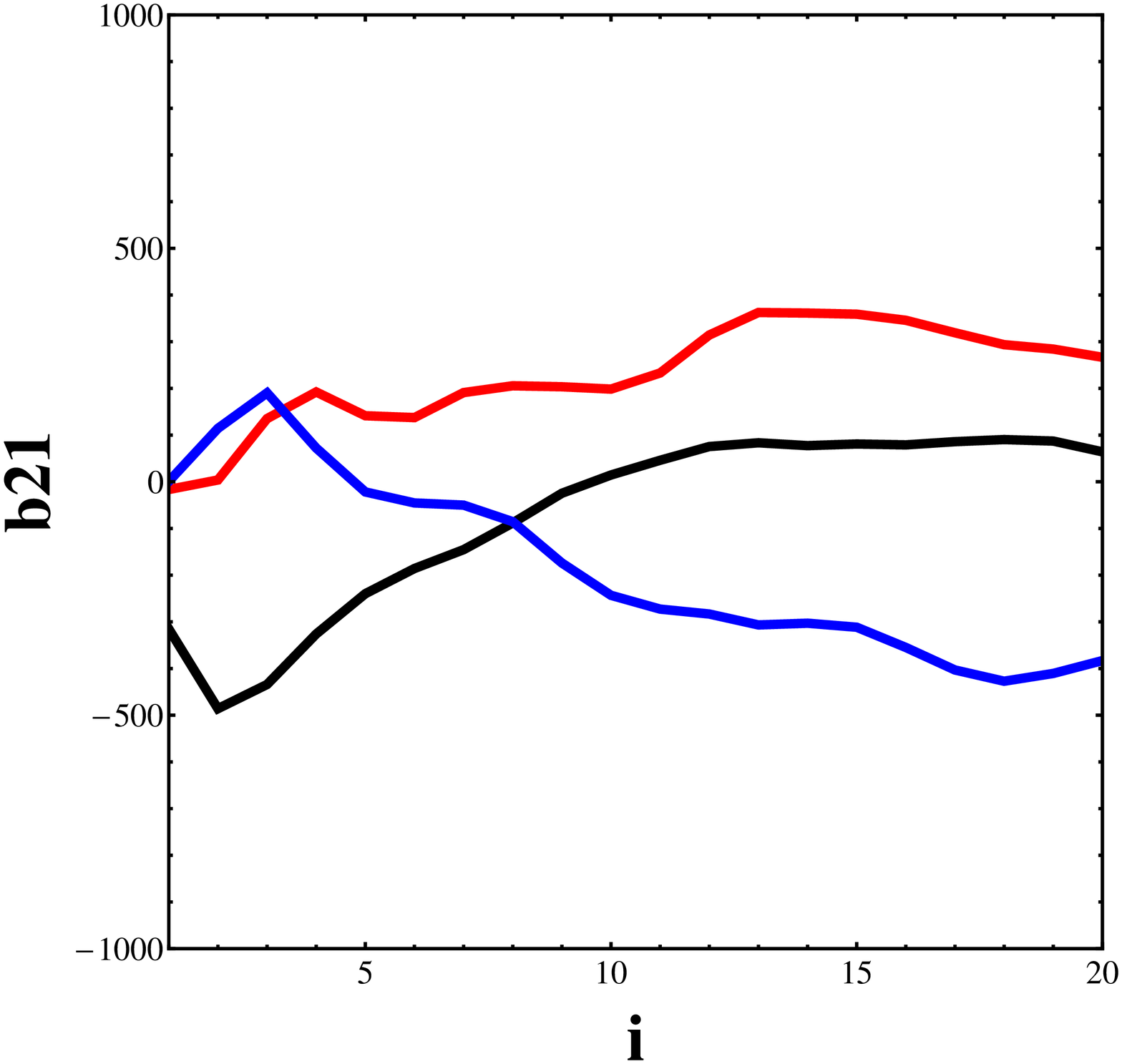}
\includegraphics[scale=0.20]{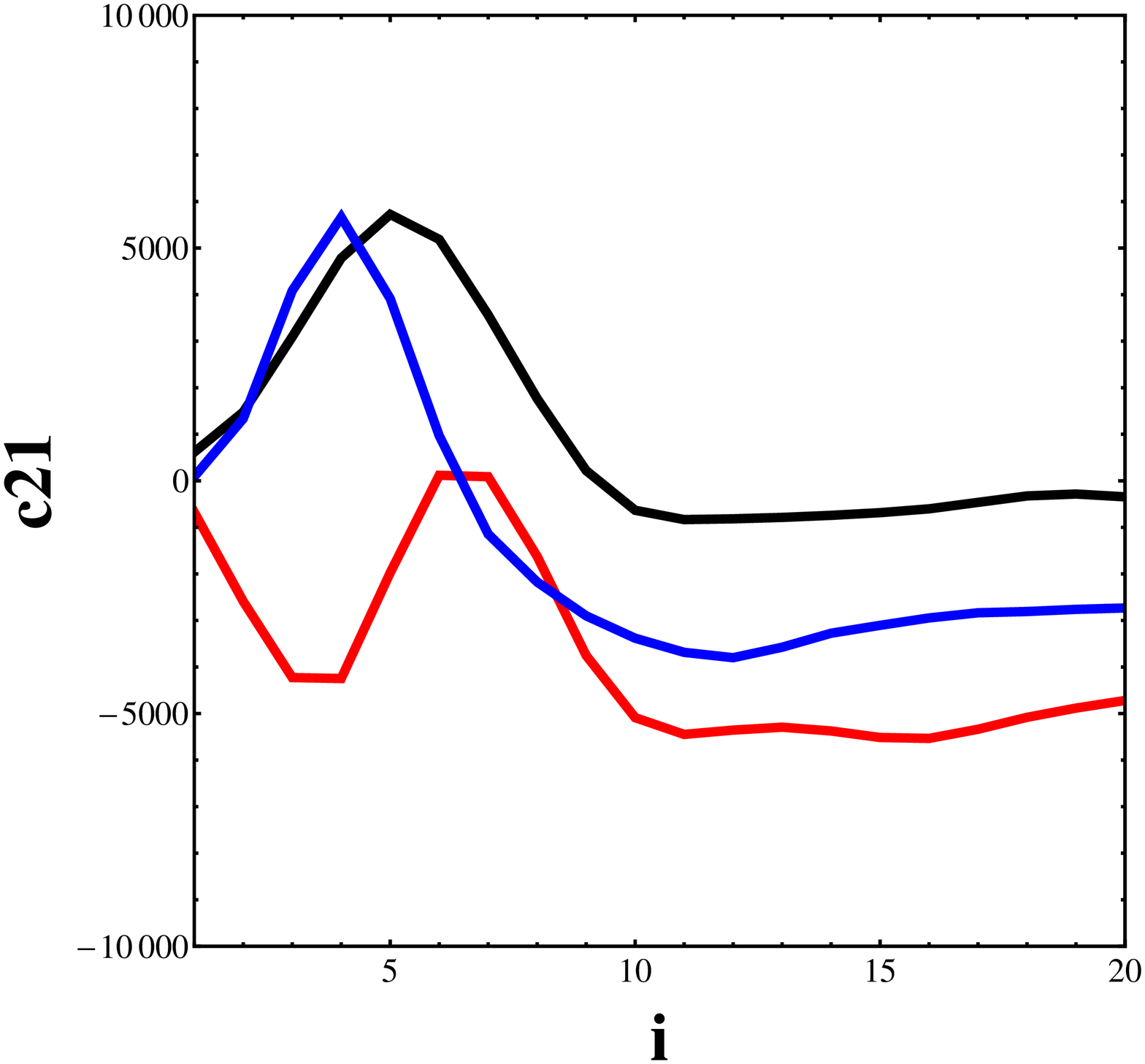}
\includegraphics[scale=0.20]{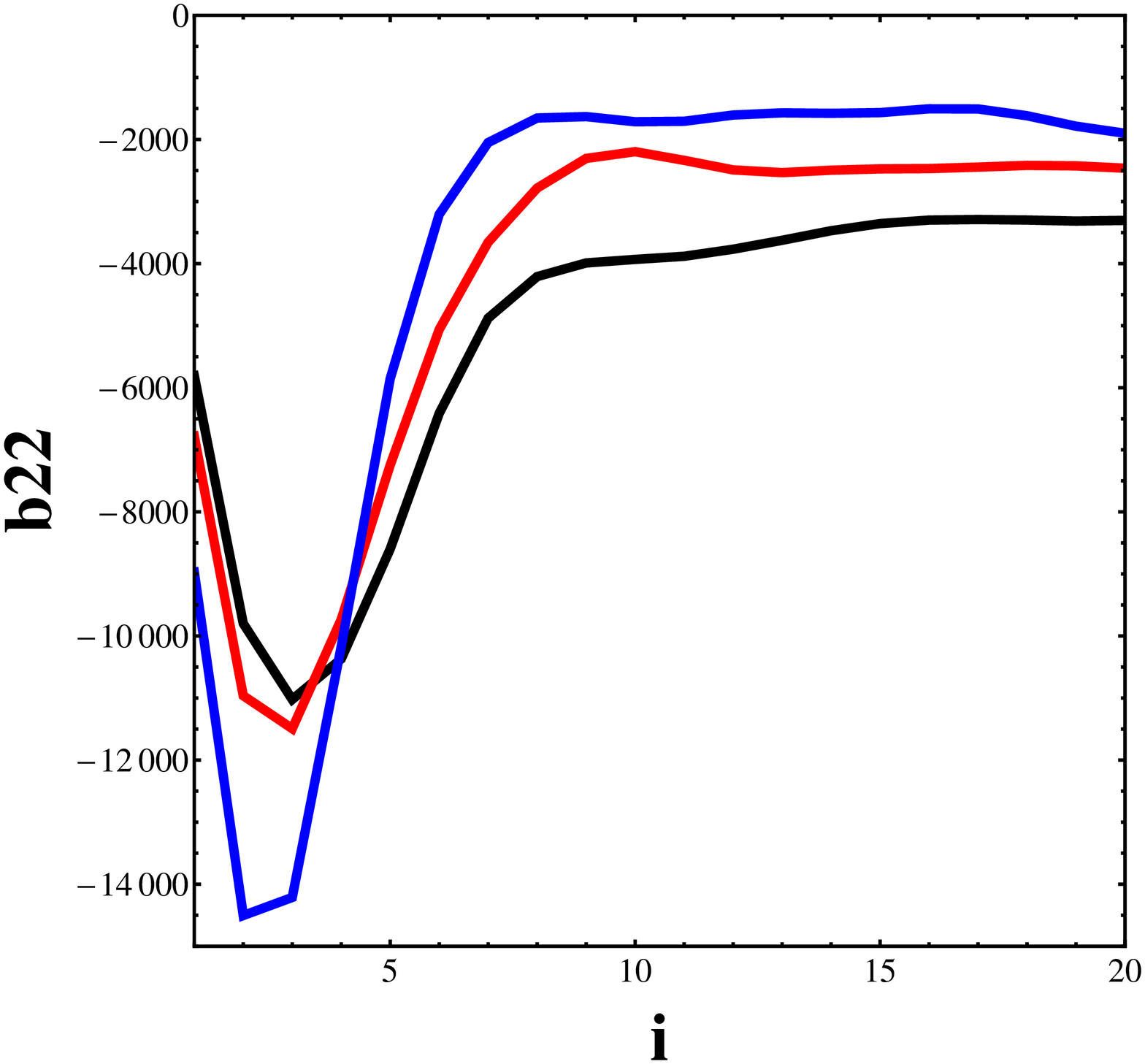}
\includegraphics[scale=0.20]{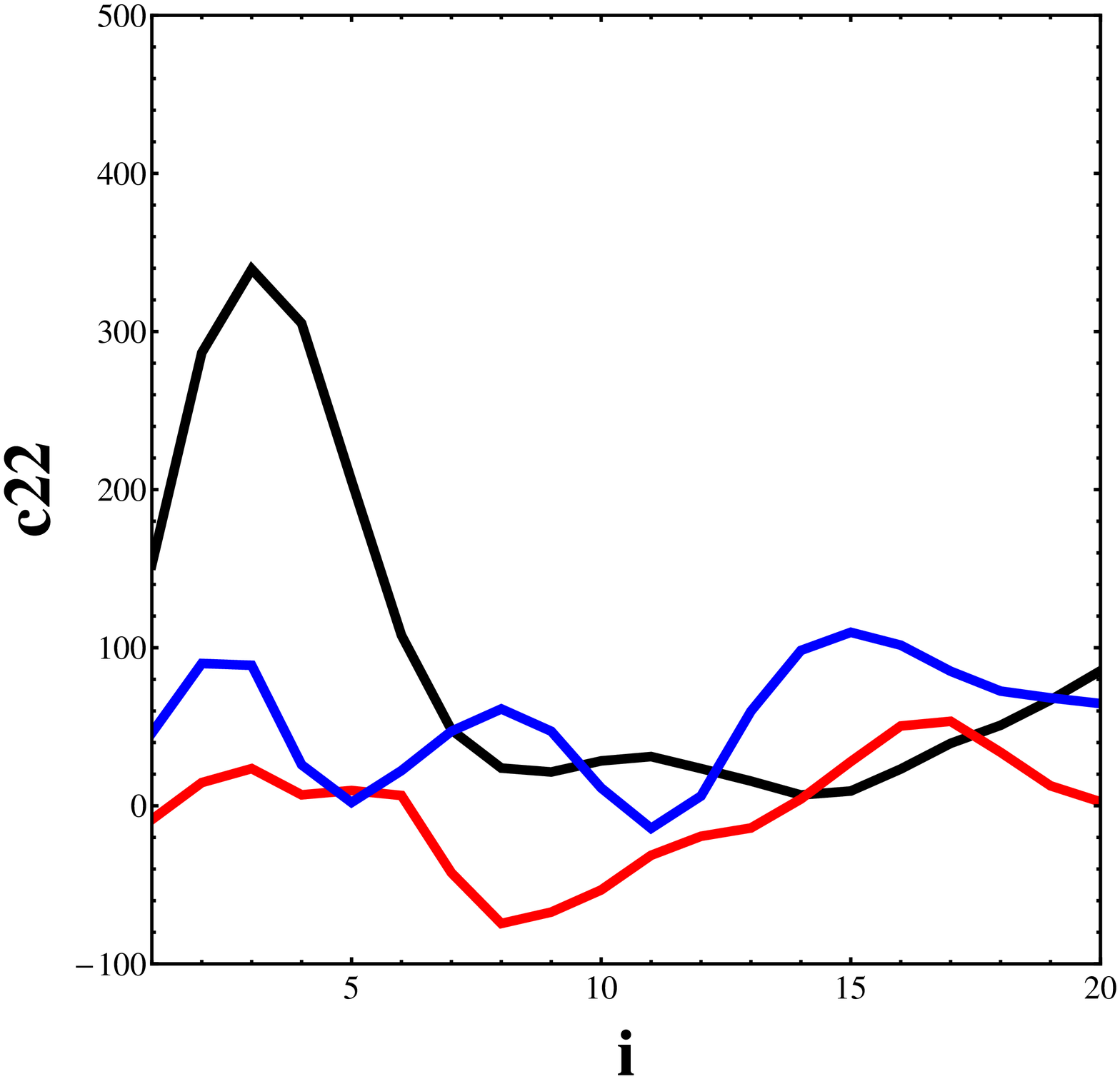}
\caption{The coefficients of the potential of Eq.(\ref{phipot})
for the three galactic models. Black (light gray in the printed
version), red(gray in the printed version) and blue (black in the
printed version) curves correspond to model $"A"$, $"B"$ and
$"C"$, respectively.} \label{figcoef}
\end{figure*}

The coefficients of Eq. (\ref{coef}) of the potential $\Phi(r,\phi)$
for the three galactic models $A$, $B$ and $C$ that are plotted in
Fig.\ref{figcoef}.

The 20 coefficients $B00$ correspond to monopole terms, the 20
coefficients $B20$ to quadrupole terms and the remaining 80
coefficients $B21,~C21,~B22,~C22$  to triaxial terms. Values are
given in the N-body units (see \citet{b33}). In the same units one
has $R=0.85$, while the corresponding pattern speeds are:
$\Omega_{pA}=5886.65$, $\Omega_{pB}=6010.36$, $\Omega_{pC}=6137.14$
(corresponding to $\approx 20-25~ km~ sec^{-1}~ Kpc^{-1}$ in
physical units). Length units in Figures 1,2,4,5,6,7b were rescaled
by the half mass radius ($R_{hm}$) of each galactic model, i.e. by a
factor of 0.1006, 0.0926 and 0.1167 for models "A", "B" and "C",
respectively.

\newpage
\section*{Appendix B: Construction of an  approximate mapping at corotation}
We show below how to construct an approximate symplectic mapping
describing the motion of stars at the corotation resonance, based on
the epicyclic approximation.

The corotation radius $r_c$ is the root for $r$ of the equation:
\begin{equation}
\Omega_p = \sqrt{F_0/r}
\end{equation}
where $F_0=\partial\Phi_0(r)/\partial r$ is the axisymmetric force.
The angular momentum at corotation is $P_{\phi c}=\Omega_pr_c^{2}$.
We define the quantities
\begin{equation}\label{newvari}
J_{\phi}=P_{\phi}-P_{\phi c},~~~~~~ \delta r=r-r_c
\end{equation}
Substituting (\ref{newvari}) in Eq.(\ref{hamrot2}), the Hamiltonian
becomes a function of the new variables
$H=H(J_{\phi},\phi,P_r,\delta r)$, which is polynomial of order 2 in
$J_\phi$. We also make a series expansion up to order 4 in $\delta
r$. Then, the Hamiltonian takes the form:
\begin{eqnarray}\label{hamstand}
H=\frac{1}{2}P^2_r+\frac{1}{2}\kappa^2_r\delta r^2
+\frac{J^2_{\phi}}{2r_c^2} -(2P_{\phi c}/r_c^3) J_{\phi} \delta r\nonumber\\
+(3P_{\phi c}/r_c^4)   J_{\phi} \delta r^2 -(1/r_c^3)J^2_{\phi}
\delta r +(3/2r_c^4)J_{\phi}^2\delta r^2
\\
+A_1 \cos (2 \phi)+A_2 \sin(2 \phi) + B_1 \delta r \cos(2 \phi) +
B_2 \delta r \sin(2 \phi)
\nonumber\\
+ C_1 \delta r^2 \cos(2 \phi) +
 C_2 \delta r^2 \sin(2 \phi) +
{\cal O} (\delta r^3)+ {\cal O}(\delta r^4)+ \ldots \nonumber
\end{eqnarray}
In Eq.(\ref{hamstand}) $\kappa_r$ is the epicyclic frequency at
corotation
\begin{equation}
\kappa_r=\sqrt{\frac{\partial^2\Phi_{eff}(r)}{\partial
r^2}}\Large{|}_{r_c}
\end{equation}
where $\Phi_{eff}=\frac{P^2_{\phi c}}{2r^2}+\Phi_0(r)$ is the
effective potential of the axisymmetric component. The constants
$A_1$, $A_2$, $B_1$, $B_2$, $C_1$, $C_2$ are computed from the
general expansion of the potential evaluated at the corotation
radius.

We now introduce a pair of epicyclic action-angle variables
$(J_r,\phi_r)$ via the relations:
\begin{equation}\label{newvar}
\delta r = \sqrt{\frac{2J_r}{\kappa_r}} \sin(\phi_r), ~~~
P_r=\sqrt{2\kappa_rJ_r} \cos(\phi_r)
\end{equation}
The lowest order terms of the Hamiltonian (\ref{hamstand}) take the
form:
\begin{eqnarray}\label{hamcor}
H = \kappa_r J_r -(2P_{\phi
c}/r_c^3)J_{\phi}\sqrt{\frac{2J_r}{\kappa_r}}\sin(\phi_r)
\nonumber\\
+\frac{J^2_{\phi}}{2r_c^2} +A_1\cos(2\phi)+A_2\sin(2\phi)+...
\end{eqnarray}
The above Hamiltonian can be `averaged' over the fast angle
$\phi_r$, i.e. the epicyclic phase. The averaging introduces a
correction of the reference radius $r_0$ around which the epicyclic
approximation is implemented, with respect to the radius of the
circular orbit $r_c$ at corotation \citep[see][p.381]{b1}. We use
the Lie method in order to make this correction via a canonical
transformation. Thus, we define the new Hamiltonian
\begin{equation}
H'=\exp(L_{X_1})H=H+L_{X_1}H+\frac{1}{2}L^2_{X_1}H+...
\end{equation}
where $L_{X_1}\equiv\{{\cdot,X_1\}}$ is the Poisson bracket
operator, and
\begin{equation}
X_1=-(2P_{\phi c}/\kappa_r r_c^3)
J_{\phi}\sqrt{\frac{2J_r}{\kappa_r}}\cos(\phi_r)
\end{equation}
The new averaged Hamiltonian has the form:
\begin{eqnarray}\label{hamave}
H'=H(\phi,J_{\phi},J_r)=\kappa_r J_r + B_1 J_r^2 +B_2 J_r J_{\phi}
+B_3 J^2_{\phi} \nonumber \\+B_4 J_r J^2_{\phi}  +B_5 \cos(2 \phi)
+B_6J_r \cos(2 \phi) +  \nonumber \\B_7 \sin(2 \phi) +B_8 J_r \sin(2
\phi)+ \ldots
\end{eqnarray}
where i) $B_s$ are known coefficients, and ii) all higher-order
terms depending on the fast angle $\phi_r$ are ignored.

In the approximation of the Hamiltonian (\ref{hamave}) $J_r$ is an
integral of motion, corresponding to a nearly constant value of the
epicyclic action along the epicyclic oscillations. On the other
hand, the variables $(\phi,J_\phi)$ yield a pendulum-like behavior,
characteristic of the corotation resonance. We will now use the
Hadjidemetriou method in order to obtain a symplectic mapping model
better describing this resonance. According to this method, the
averaged Hamiltonian $H'(J_r,\phi,J_{\phi})$ is employed in order to
define a generating function $S$ of the second kind:
\begin{equation}
S=\phi J'_{\phi}+T_r H'(J_r,\phi,J'_{\phi})
\end{equation}
where $T_r=2\pi/\kappa_r$ is the epicyclic period. The symplectic
mapping equations are then given by:
\begin{eqnarray}
~~~~~~~~~~~~~~~~~~~~~J_{\phi}=\frac{\partial S}{\partial
\phi}=G(\phi,J_r,J'_{\phi})\nonumber
\\\phi'=\frac{\partial S}{\partial J'_{\phi}}=F(\phi,J_r,J'_{\phi})
\end{eqnarray}
Solved for $\phi'$, $J_\phi'$, these equations give the mapping
$(\phi,J_\phi)\rightarrow(\phi',J_\phi')$ after one epicyclic
period. The variable $J_r$ is a constant parameter of the mapping.
In particular, the value $J_r=0$ corresponds to orbits with a zero
epicyclic oscillations, i.e., asymptotic to the Lagrangian points
$L_1$, $L_2$, while for $J_r\neq 0$ we find orbits asymptotic to the
short period orbits $PL_1$ or $PL_2$.

Setting $J_r$=0 we have the expressions of $J'_{\phi}$ and $\phi'$:
\begin{eqnarray}\label{newmap}
~~~~~~~~~~~J'_{\phi}=J_{\phi} + C_1 \cos(2 \phi) +C_2 \sin(2 \phi )
\nonumber\\
\phi'= \phi+ C_3 J_{\phi} +C_4 \cos(2\phi) +C_5 \sin(2\phi )
\end{eqnarray}
with $C_1,... C_5$ known coefficients.

It is straightforward to show that the mapping (\ref{newmap}) takes
the form of the well known Standard map \citep{b46} after some
appropriate transformations which include the following: (a)
Eliminate the $\cos(2\phi)$ term, (b) eliminate the factor 2 inside
the $\sin$ term  and (c) eliminate the coefficient $C_3$ of the
$J_{\phi}$ term.

Non-zero coefficients $C_1,~C_4$ indicate that the main axes of the
bar of the galaxy are not aligned with the axes $(x,~y)$ of the
coordinate system. We find the bar's axis angular position by
calculating the coordinates of the main periodic orbits of the
mapping (\ref{newmap}). The system of equations:

\begin{equation}
J'_{\phi} = J_{\phi},~~~ \phi' =\phi
\end{equation}
gives the solution $J_{\phi}=0$ and $\phi=\pi/2+\delta \phi$. Making
for some constant $\delta \phi$ the transformation $\phi\rightarrow
\theta+\delta \phi$ the mapping (\ref{newmap}) takes the form:

\begin{eqnarray}
~~~~~~~~~~~~~~~x_v=C'_1 J_{\phi} + \theta + C'_2
\sin(2\theta)\nonumber
\\y_v=J_{\phi} +C'_3\sin(2 \theta)
\end{eqnarray}
Finally  making the transformation $\theta\rightarrow \Theta/2$ and
$J_{\phi}\rightarrow Y/(2C'_1)$ we arrive at the standard map:

\begin{eqnarray}\label{newstand}
~~~~~~~~~~~~~~~~~\Theta'=\Theta + Y+ K \sin(\Theta)\nonumber
\\Y'=Y +K \sin(\Theta)
\end{eqnarray}
 $K$ is a non linearity parameter depending on the non-axisymmetric perturbation
of the galactic model.

One way to quantify  how good is the mapping approximation
(\ref{newstand}) is by comparing the eigenvalues of its unstable
periodic orbits with the ones derived from the original Hamiltonian.

Taylor expanding the Hamiltonian (\ref{hamave}) up to second order
in the angle $\phi$ around $\delta \phi$ we find the approximative
Hamiltonian:

\begin{equation}
H\approx\kappa_r J_r + D_1 J^2_{\phi} +2\omega^2_0 \phi^2
\end{equation}
with $D_1$ and $\omega_0$ known coefficients.

We then obtain a second-order differential equation for the angle
$\phi$:

\begin{equation}
\ddot{\phi}-8D_1\omega^2_0 \phi=0
\end{equation}
with solution: $\phi(t)= A e^{\pm 2\omega_0\sqrt{ 2D_1} t}$.

The unstable eigenvalue  on an apocentric Poincar\'{e} map is
$\lambda=4 \pi \omega_0\sqrt{ 2D_1}/\kappa_r$. This must be compared
with the eigenvalue derived from the monodromy matrix of the map of
Eq. \ref{newstand} for each galactic model. Table I below shows this
comparison.

\begin{center}
\begin{tabular}{|c|}
\hline
 Table I. ~~~~  Comparison of the eigenvalues \\
\end{tabular}
\begin{tabular}{|c|c|c|}
  \hline
Galactic Model & $\lambda~ Hamiltonian$ & $\lambda~ Mapping \ref{newstand}$ \\
\hline
$ A$ & $4.465$ & $4.461$ \\
  \hline
 $ B$ & $8.441$ & $6.450$ \\
  \hline
$C$ & $20.736$ & $11.189$ \\

  \hline
\end{tabular}
\end{center}

Form Table I we find that the mapping \ref{newstand} is a good
approximation in the case of model "A" while we have the largest
deviation in the case of model "C".

Thus the mapping \ref{newstand} provides only qualitative results in
the case of models "B" and "C".

 On the other hand, the use of a mapping model
is motivated by the fact that the analysis of the Moser domain is
greatly facilitated in such mappings using the same method as in
\citet{b18}. Briefly, the steps are the following: (a) we make a
Taylor expansion of the mapping (\ref{newstand}) around the
hyperbolic point $(0,0)$ up to a desired order, (b) we introduce a
new linear symplectic transformation $(\Theta,Y)\rightarrow(u,v)$,
and finally (c) we find the integrals of motion that correspond to
the Moser invariant curves, which are hyperbolas $c=\xi\eta$ in some
new variables $(\xi,\eta)$, via convergent series $\Phi(\xi,\eta)$.
The procedure is described, in detail, in section 4 of \citet{b18}
and the method of calculating the transformation $\Phi$ is described
by \citet{b8}.

In order to find the limits of the region of convergence, inside
which these analytical convergent series exist, we use the
d'Alembert criterion that determines the convergence radius along
various directions with angles $\phi=\tan^{-1}(\eta/\xi)$ in the
plane of the new variables $(\xi,\eta)$.  The limiting value of $c$
for each angle is given by the relation:
\begin{equation}
c = \rho_c^2\cos(\phi)\sin(\phi)
\end{equation}

We find first the Moser region of convergence on the ($\xi,\eta$)
plane of the new variables, which is the region in the four
quadrants around the origin limited by the hyperbolas with
$c=c_{lim}$.  In order to convert this region to the old variables
$(\Theta,Y)$ of the mapping (\ref{newstand}) we place points on a
grid of hyperbolas. In each quadrant the distribution of points is
found inside the limiting hyperbola $c=|c_{lim}|=\xi\eta$. The first
point $A$ on every hyperbola is taken on the diagonal $\xi=\eta$,
i.e. $\xi_0=\eta_0=\sqrt{c}$ and  the last point $B$ on every
hyperbola must be the image of $A$ under the mapping:
 \begin{eqnarray}
\xi'&=&\Lambda(c)\xi=(\lambda_1+w_2c+w_3c^2 +...)\xi \nonumber
\\ \eta'&=&\frac{1}{\Lambda(c)}\eta=(\lambda_2+q_2c+q_3 c^2+...)\eta
\end{eqnarray}

These regions correspond to the unstable direction of the
corresponding hyperbolic point.

Then by making the back transformation to the old variables of the
mapping (\ref{newstand}) we have the same region of convergence on
the $(\Theta,Y)$ plane  and finally we make the transformation to
the variables original $(x,y)$ of the configuration space of the
galactic models. Hence we produce the images of the Moser domain in
the phase space ($\Theta,Y$) or the configuration space $(x,y)$ as
in Fig. 7.


\end{document}